\documentclass{aa}
\usepackage{txfonts}
\usepackage{graphicx}
\usepackage{caption}
\usepackage{subcaption}
\usepackage{color}
\usepackage{tablefootnote}

\usepackage{tikz}
\usetikzlibrary{arrows,shapes,backgrounds}

\begin{document} 
\titlerunning{}
\authorrunning{Mountrichas et al.}
\titlerunning{The role of AGN and its obscuration on the position of the host galaxy relative to Main Sequence}

\title{The role of AGN and its obscuration on the position of the host galaxy relative to Main Sequence}

\author{G. Mountrichas\inst{1}, V. Buat\inst{2,3}, G. Yang\inst{4,5}, M. Boquien\inst{6}, D. Burgarella\inst{2}, L. Ciesla\inst{2}, K. Malek\inst{7,2}, R. Shirley\inst{8,9}}
          
    \institute {Instituto de Fisica de Cantabria (CSIC-Universidad de Cantabria), Avenida de los Castros, 39005 Santander, Spain
              \email{gmountrichas@gmail.com}
           \and
             Aix Marseille Univ, CNRS, CNES, LAM Marseille, France. 
                \email{ veronique.buat@lam.fr}  
              \and
                 Institut Universitaire de France (IUF)
                \and
                Department of Physics and Astronomy, Texas A\&M University, College Station, TX 77843-4242, USA 
               \and
                George P. and Cynthia Woods Mitchell Institute for Fundamental Physics and Astronomy, Texas A\&M University, College Station, TX 77843-4242, USA 
                \and
                Centro de Astronom\'ia (CITEVA), Universidad de Antofagasta, Avenida Angamos 601, Antofagasta, Chile
          \and
		National Centre for Nuclear Research, ul. Hoza 69, 00-681 Warszawa, Poland
     \and
	Astronomy Centre, Department of Physics \& Astronomy, University of Southampton, Southampton, SO17 1BJ, UK 
	 \and
    Institute of Astronomy, University of Cambridge, Madingley Road,
  Cambridge CB3 0HA, UK}

\abstract {We use X-ray Active Galactic Nuclei (AGN) observed by the {\it{Chandra}} X-ray Observatory within the 9.3\,deg$^2$ Bo$\rm \ddot{o}$tes field of the NDWFS to study whether there is a correlation between X-ray luminosity (L$_X$) and star formation rate (SFR) of the host galaxy, at $\rm 0.5<z<2.0$, with respect to the position of the galaxy to the main sequence (SFR$_{norm}$). About half of the sources in the X-ray sample have spectroscopic redshifts. We also construct a reference galaxy catalogue. For both datasets, we use photometric data from optical to the far infrared, compiled by the HELP project and apply spectral energy distribution (SED) fitting, using the X-CIGALE code. We exclude quiescent sources from both the X-ray and the reference samples. We also account for the mass completeness of our dataset, in different redshifts bins. Our analysis highlights the importance of studying the  SFR-L$_X$ relation, in a uniform manner, taking into account the systematics and selection effects. Our results suggest that, in less massive galaxies ($\rm log\,[M_*(M_\odot)] \sim 11$), AGN enhances the SFR of the host galaxy by $\sim 50\%$ compared to non AGN systems. A flat relation is observed for the most massive galaxies. SFR$_{norm}$ does not evolve with redshift. The results, although tentative, are consistent with a scenario in which, in less massive systems, both AGN and star formation (SF) are fed by cold gas, supplied by a merger event. In more massive galaxies, the flat relation could be explained by a different SMBH fuelling mechanism that is decoupled from the star formation of the host galaxy (e.g. hot diffuse gas). Finally, we compare the host galaxy properties of X-ray absorbed and unabsorbed sources. Our results show no difference which suggests that X-ray absorption is not linked with the properties of the galaxy.} 

   
\maketitle

\section{Introduction}

The last years we have witnessed a significant progress in our understanding of how supermassive black holes (SMBH) form and grow over cosmic time. It is also widely accepted and well established that there is a correlation between the mass of the SMBH and the properties of the galactic bulge \citep[e.g.][]{Magorrian1998, Ferrarese2000}. However, it is still not clear what are the physical mechanisms that drive the black hole (BH) growth and if there is a connection between the properties of BH (e.g. accretion luminosity) and those of the host galaxy (e.g., star formation rate (SFR), stellar mass). 

BHs grow through accretion of cold gas onto the accretion disk. This gas originates from at least nine orders of magnitude larger scales, either from the host galaxy or the extragalactic environment. Various mechanisms have been suggested to drive the gas from kpc to sub-pc scales \citep[for a review see][]{Alexander2012}. The SMBH becomes active when material is accreted onto it and then the galaxy is called an Active Galactic Nuclei (AGN). Star formation (SF) is also driven by cold gas. Moreover, AGN activity and SF peak at the same cosmic time \citep[$\rm z \sim 2; e.g.,$][]{Boyle1998, Boyle2000, Sobral2013}. These advocate for a connection between the BH activity and galaxy growth. However, the nature of the connection, if any, is still a matter of debate.

Hydrodynamical simulations and semi-analytic models require negative AGN feedback to suppress the formation of stars and avoid the production of too many massive galaxies. Cooling outflows generated by strong winds which are produced by the AGN \citep[e.g.][]{Debuhr2012} remove or heat the star forming gas. At high accretion rates, this feedback is radiation-based ("quasar mode") whereas at lower accretion rates, AGN feedback is provided mechanically by the kinetic energy of jets \citep{Scheuer1974, Blandford1974}. Nevertheless, AGN feedback can work both ways \citep[e.g.][]{Zinn2013}. In the late gas poor phase, AGN feedback may quench star formation. In the gas rich phase, though, AGN outflows can provide positive feedback to their hosts, by over-compressing cold gas \citep[e.g.][]{Zubovas2013}.

Observationally, a popular method to examine whether and how AGN affect the evolution of their host galaxy is to examine if there is a link between the SFR of the galaxy and the X-ray luminosity of the AGN. The former is an indicator of the galaxy growth while the latter is a proxy of the AGN power. Results from the first studies that examined these two properties had controversial results \citep[e.g.][]{Lutz2010, Page2012}. These discrepancies were attributed to e.g. selection effects and low number statistics as a result of the small sample sizes available \citep[e.g.][]{Harrison2012, Rosario2012}. Different conclusions may also be drawn, depending on the binned parameter \citep{Hickox2014, Lanzuisi2017, Brown2019, Masoura2021}. This was attributed to AGN variability that is significantly shorter compared to the average timescales of star formation \citep[e.g.][]{Hickox2014}.

It is well known that star forming galaxies follow a tight correlation between their SFR and stellar mass, known as main sequence (MS) of star formation \citep[e.g.,][]{Noeske2007, Elbaz2007, Pannella2009}. Therefore, more recent studies started to explore the link between SFR and L$_X$ relative to the MS. This allows the study of the two properties taking into account the stellar mass and redshift of the host galaxy, i.e. the position of the host galaxy with respect to the MS. The parameter utilized for that purpose was the SFR$_{norm}$, defined as the ratio of the SFR of X-ray AGN to the SFR of star forming MS galaxies. The first results following this more refined approach showed that AGN and star forming galaxies have similar mean SFR$_{norm}$ values, although the SFR distributions of the two populations are different. They also find no evolution of the SFR$_{norm}$ with redshift \citep{Mullaney2015}. Later works, found that the distribution of SFR$_{norm}$ of higher L$_X$ AGN is narrower and closer to that of MS galaxies than that of lower L$_X$ \citep{Bernhard2019}. However, the results of these studies were limited by the small sample sizes \citep{Mullaney2015} or the narrow redshift range and thus X-ray luminosity baselines spanned by the datasets \citep{Bernhard2019}. Deploying a significantly larger sample, with more than $3000$ X-ray sources, revealed a non flat relation between the SFR of the host galaxy and the AGN power \citep{Masoura2018} and a hint of evolution of SFR$_{norm}$ with redshift \citep{Masoura2021}.  Nevertheless, a downside of these studies is that they did not define their own reference (non AGN) galaxy sample to estimate SFR$_{norm}$. Instead, they used relations from the literature to estimate the star formation of main sequence galaxies \citep[e.g. equation 9 in][]{Schreiber2015} with which they compare the SFR of X-ray AGN. This approach hints at a number of systematics. For example, different methods are utilized to estimate galaxy properties, there is no exact definition of MS and different selection criteria are applied on X-ray and non X-ray galaxy samples in different studies.

\cite{Shimizu2015} compared the SFR of X-ray AGN in the local universe ($\rm z<0.1$) with that of MS galaxies, by constructing their own MS galaxy sample and using the same methods to measure the SFR and stellar mass of sources in both datasets. They use different datasets to estimate galaxy properties for AGN and non AGN systems. Based on their results, X-ray luminous AGN have decreased SFR compared to non AGN systems. \cite{Santini2012} compared the average SFR of X-ray AGN with inactive galaxies of similar mass. Their analysis showed that lower luminosity AGN (L$_X\leq 10^{43.5}$\,erg\,s$^{-1}$) have enhanced star formation (by $\sim 0.26$\,dex) compared to non AGN systems, while for higher luminosity AGN the enhancement is more evident ($\sim 0.56$\,dex). \cite{Rosario2013} found that the mean SFR of broad line AGN are consistent with those of normal massive star forming galaxies. Recently, \cite{Florez2020} used X-ray selected AGN in the Stripe 82 field and compared their SFR with non X-ray galaxies, in a wide range of redshift and luminosities. They estimated SFR for both samples in a consistent way, by applying SED fitting on both datasets. Their analysis showed that X-ray AGN have on average 3-10$\times$ higher SFR than their control galaxy sample, at the same stellar mass and redshift.

Another aspect of the AGN-galaxy co-evolution is whether the absorption of the system is linked with the host galaxy properties. According to the simple unification model \citep[e.g.][]{Antonucci1993}, the classification of AGN into absorbed (type 2) and unabsorbed (type 1) is only a geometrical effect. Type 1 refer to the face on and type 2 to edge-on AGN. This simplified picture has been updated in more complex scenarios within the unification framework. However, the inclination determines the classification even in these updated unified pictures of the AGN structure \citep[e.g.][]{Ogawa2021}. On the other hand, according to the evolutionary model \citep[e.g.][]{Ciotti1997, Hopkins2006} AGN growth coincides with host galaxy activity, during which the AGN appears absorbed. At later stages, as the AGN becomes more powerful it blows away the SF material and thus SF is quenched. Therefore, comparison of  host galaxy properties between the two AGN types could favour one over the other model \citep[e.g.][]{Merloni2014, Chen2015, Zou2019, Masoura2021}. However, these efforts are hindered by the fact that absorption may occur, not only close to the accretion disk, but also on galactic scales \citep[e.g.][]{Maiolino1995, Circosta2019, Malizia2020}. Furthermore, classification criteria based on different wavelengths, e.g. X-ray criteria, optical spectra, optical/mid-IR colours, are sensitive to different levels of absorption and this may lead to different characterization of a source \citep[e.g.][]{Li2019, Masoura2020, Ogawa2021}.

In this paper, we revisit the SFR-L$_X$ relation using X-ray sources from the 9.3\,deg$^2$ Bo$\rm \ddot{o}$tes field. In Section 2, we describe the datasets and the available photometry. The SED fitting analysis, the mass completeness of our samples and the reliability of the host galaxy measurements are presented in Section 3. In Section 4, we describe how we identify and exclude quiescent systems from our datasets and present the systematics introduced by different definitions of the star forming MS. The relation between SFR$_{norm}$ and L$_X$ is studied in Section 5. In the last Section we compare host galaxy properties of X-ray absorbed and unabsorbed sources. The classification is parametererized using the hydrogen column density, $\rm N_H$.

%To facilitate comparison with previous studies, we measure the SFR$_{norm}$ parameter using for the estimation of the SFR of MS galaxies the same expression from the literature that was used in previous similar works. We also construct our own MS galaxy sample, applying the same photometric selection criteria used on the X-ray sample. The same SED fitting analysis is performed on both datasets. We exclude from the MS sample X-ray sources and systems with a strong AGN component. Finally, quiescent galaxies are removed from both catalogues. This allows us to re-estimate SFR$_{norm}$ in a uniform manner, with better control of systematics and selection biases that affect previous similar studies. We also compare host galaxy properties of X-ray absorbed and unabsorbed sources. The classification is parametererized using the hydrogen column density, $\rm N_H$.

Throughout this work, we assume a flat $\Lambda$CDM cosmology with $H_ 0=69.3$\,Km\,s$^{-1}$\,Mpc$^{-1}$ and $\Omega _ M=0.286$.

\section{Data}
\label{sec_data}

In this work, we use X-ray AGN observed by the {\it{Chandra}} X-ray Observatory within the 9.3\,deg$^2$ Bo$\rm \ddot{o}$tes field of the NOAO Deep Wide-Field Survey (NDWFS). The catalogue is compiled and fully described in \cite{Masini2020}. It consists of 6891 X-ray point sources with an exposure time of about 10\,ks per XMM pointing and a limiting flux of $4.7\times 10^{-16}$, $1.5\times 10^{-16}$ and $\rm 9\times 10^{-16}\,erg\,cm^{-2}\,s^{-1}$, in the $\rm 0.5-7\,keV$, $\rm 0.5-2\,keV$ and $\rm 2-7\,keV$ energy bands, respectively. 2346 ($\sim 33\%$) of the X-ray sources in this catalogue, have available spectroscopic redshifts (specz). Specz are obtained by cross-matching the $I-$band of the NDWFS catalogue with the AGES catalogue \citep{Kochanek2012}, using a matching radius of 0.5 arcsec. We use these specz sources to optimize the spectral energy distribution (SED) grid used in our analysis (see Section \ref{sec_analysis}). For the remaining sources, we use hybrid photometric redshifts \citep[photoz;][]{Duncan2018a, Duncan2018b, Duncan2019} that are available in the \cite{Masini2020} catalogue. A Gaussian process redshift code, GPz, is combined with template photoz estimates through hierarchical Bayesian combination and produce a hybrid estimate that significantly improves the individual methods. The normalized median absolute deviation of the photoz is $\sigma _{NMAD}=0.054$ and the fraction of outliers, defined as $\rm |photoz-specz|/(1+specz)>0.15$,  is $13.23\%$. The X-ray absorption of each X-ray AGN is available and is parameterized with $\rm N_H$. $\rm N_H$ has been calculated from Hardness Ratio (HR) estimations which are measured using the Bayesian Estimator for Hardness Ratio \citep[BEHR;][]{Park2006}. A fixed Galactic absorption of $\rm N_{H, Gal}=1.04\times 10^{20}\,cm^{-2}$ is assumed. The average uncertainty on the $\rm N_H$ values is $\sim 17\%$.

The X-ray catalogue is cross-matched with the Bo$\rm \ddot{o}$tes photometric catalogue produced by the HELP\footnote{The {\it Herschel} Extragalactic Legacy Project (HELP, http://herschel.sussex.ac.uk/) is a European funded project to analyse all the cosmological fields observed with the {\it Herschel} satellite. All the HELP data products can be accessed on HeDaM (http://hedam.lam.fr/HELP/)} collaboration. HELP provides homogeneous and calibrated multiwavelength data over  the {\it Herschel}  Multitiered Extragalactic Survey \citep[HerMES,][]{Oliver2012a}  and the  H-ATLAS survey \citep{Eales2010} from the ultraviolet (UV) to near-infrared (NIR). The position of NIR/IRAC sources are then used as prior information to extract sources in the {\it{Herschel}} maps. IRAC1 positions are used for the Bo$\rm \ddot{o}$tes field \citep{Shirley2019}, where IRAC1 is the [3.6]\,$\mu$m bands of {\it{Spitzer}}. The  XID+ tool \citep{Hurley2017}, developed for this purpose, uses a  Bayesian probabilistic framework  and works with prior positions. Fluxes are measured for the Spitzer MIPS/24 microns, and Herschel PACS and SPIRE bands. In this work, only the MIPS and SPIRE fluxes are considered, given the much lower sensitivity of the PACS observations for this field \citep{Oliver2012}.

The cross-match of the X-ray sample with the HELP dataset is done using 1'' radius and the I-band coordinates provided in the X-ray catalogue. This process results in 5887 matches. Since, in our analysis, we need to accurately measure the host galaxy properties via SED fitting, we require the best possible photometric coverage for the sources while at the same time keep the size of the dataset large. For this reason, we restrict the X-ray sample to those sources that have been detected in the following photometric bands: {\it{u, g or B, R, I, z, H or K$_s$}}, IRAC1, IRAC2 and MIPS/24, where IRAC2 is the [4.5]\,$\mu$m band of {\it{Spitzer}}. This reduces X-ray sources to 2778. Since our dataset does not have  UV data, which can directly trace the young stellar population, we restrict our sample to sources that lie at redshift higher than 0.5. At $\rm z>0.5$, the {\it{u}} band is redshifted to rest-frame wavelength $<2000\,\AA$, allowing observation of the emitted radiation from young stars. This further reduces the X-ray sources to 2338. Nearly half of the sources have available {\it{Herschel}} photometry (see Section \ref{sec_reliability}).

In the first part of our analysis, we compare the SFR of X-ray AGN with the SFR of normal, i.e. non-AGN, star forming systems. For that purpose, we apply the same SED fitting analysis in both the X-ray catalogue and to a comparison galaxy sample (reference galaxy catalogue hereafter). This enables us to perform a fully consistent comparison between the SFR of X-ray AGN with that of non X-ray systems. The reference galaxy catalogue is provided by the HELP project. The same photometric criteria are applied, for homogenisation with the X-ray catalogue. We also exclude, from the reference galaxy catalogue, sources with X-ray emission, i.e. sources that are included in the X-ray sample. This results in 56,627 sources at $\rm z>0.5$. Next, we re-estimate SFR$_{norm}$ using the expression (9) of \cite{Schreiber2015}, that parametrizes the SFR of main sequence galaxies. This allows us to facilitate a fair comparison with previous X-ray AGN studies, that used the same expression \citep[e.g.][]{Mullaney2015, Masoura2018, Bernhard2019}. Most importantly, it allows us to examine how the systematics and selection effects of the latter methodology affect the correlation of SFR$_{norm}$ with L$_X$.

\begin{table*}
\caption{The models and the values for their free parameters used by X-CIGALE for the SED fitting of our galaxy sample. For the definition of the various parameter see section \ref{sec_cigale}.} 
\centering
\setlength{\tabcolsep}{1.mm}
\begin{tabular}{cc}
       \hline
Parameter &  Model/values \\
	\hline
\multicolumn{2}{c}{Star formation history: delayed model and recent burst} \\
Age of the main population & 1500, 2000, 3000, 4000, 5000 Myr \\
e-folding time & 200, 500, 700, 1000, 2000, 3000, 4000, 5000 Myr \\ 
Age of the burst & 50 Myr \\
Burst stellar mass fraction & 0.0, 0.005, 0.01, 0.015, 0.02, 0.05, 0.10, 0.15, 0.18, 0.20 \\
\hline
\multicolumn{2}{c}{Simple Stellar population: Bruzual \& Charlot (2003)} \\
Initial Mass Function & Chabrier (2003)\\
Metallicity & 0.02 (Solar) \\
\hline
\multicolumn{2}{c}{Galactic dust extinction} \\
Dust attenuation law & Charlot \& Fall (2000) law   \\
V-band attenuation $A_V$ & 0.2, 0.3, 0.4, 0.5, 0.6, 0.7, 0.8, 0.9, 1, 1.5, 2, 2.5, 3, 3.5, 4 \\ 
\hline
\multicolumn{2}{c}{Galactic dust emission: Dale et al. (2014)} \\
$\alpha$ slope in $dM_{dust}\propto U^{-\alpha}dU$ & 2.0 \\
\hline
\multicolumn{2}{c}{AGN module: SKIRTOR)} \\
Torus optical depth at 9.7 microns $\tau _{9.7}$ & 3.0, 7.0 \\
Torus density radial parameter p ($\rho \propto r^{-p}e^{-q|cos(\theta)|}$) & 1.0 \\
Torus density angular parameter q ($\rho \propto r^{-p}e^{-q|cos(\theta)|}$) & 1.0 \\
Angle between the equatorial plan and edge of the torus & $40^{\circ}$ \\
Ratio of the maximum to minimum radii of the torus & 20 \\
Viewing angle  & $30^{\circ}\,\,\rm{(type\,\,1)},70^{\circ}\,\,\rm{(type\,\,2)}$ \\
AGN fraction & 0.0, 0.1, 0.2, 0.3, 0.4, 0.5, 0.6, 0.7, 0.8, 0.9, 0.99 \\
Extinction law of polar dust & SMC \\
$E(B-V)$ of polar dust & 0.0, 0.2, 0.4 \\
Temperature of polar dust (K) & 100 \\
Emissivity of polar dust & 1.6 \\
\hline
\multicolumn{2}{c}{X-ray module} \\
AGN photon index $\Gamma$ & 1.8 \\
Maximum deviation from the $\alpha _{ox}-L_{2500 \AA}$ relation & 0.2 \\
LMXB photon index & 1.56 \\
HMXB photon index & 2.0 \\
\hline
Total number of models (X-ray/reference galaxy catalogue) & 235,224,000/22,968,000 \\
\hline
\label{table_cigale}
\end{tabular}
\end{table*}

\section{Analysis}
\label{sec_analysis}

In this Section, we describe the SED fitting analysis we perform to measure galaxy properties. We define the mass completeness of the data in different redshift ranges and describe the final selection criteria we apply on the datasets. Finally, we examine the reliability of the SED fitting measurements.

\subsection{X-CIGALE}
\label{sec_cigale}

To measure the properties of our selected galaxies, we perform SED fitting using the X-CIGALE code \citep{Yang2020}. X-CIGALE is a new branch of the CIGALE fitting algorithm \citep{Boquien2019} that adds the ability to model the X-ray emission of galaxies and the presence of polar dust. The latter, accounts for extinction of the UV and optical emission in the poles of AGN. The re-emitted radiation is considered isotropic, thus contributes to the infrared (IR) emission of both type 1 and 2 AGN. The improvements that these new features add in the results of the fitting process are described in \cite{Yang2020} and \cite{Mountrichas2021}. 

The galaxy component is built using a delayed star formation history (SFH) with the functional form SFR$\propto t\times \exp(-t/{\tau})$. A star formation burst is also included and modelled as a constant ongoing star formation no longer than 50\,Myr. The burst is  superimposed to the delayed SFH \citep{Buat2019}. Stellar emission is modelled using the \cite{Bruzual_Charlot2003} single stellar populations template. The initial mass function (IMF) of \cite{Chabrier2003} is adopted with metallicity equal to the solar value (0.02). Stellar emission is attenuated following \cite{Charlot_Fall_2000}. The IR SED of the dust heated by stars is implemented with the \cite{Dale2014} model. In this model the star-forming component is parametrised by a single parameter $\alpha$ definded as $dM_d (U) \propto U^{-\alpha}dU$, with $M_d$ being the dust mass and $U$ the radiation field intensity. AGN emission is modelled using the SKIRTOR templates \citep{Stalevski2012, Stalevski2016}. A detailed description of the SKIRTOR implementation in (X-) CIGALE is given in \cite{Yang2020}. The AGN fraction, $\rm frac_{AGN}$, is defined as the ratio of the AGN IR emission to the total IR emission of the galaxy. A polar dust component ($E_{B-V}$) is added and modelled as a dust screen absorption and a grey-body emission. The Small Magellanic Cloud extinction curve \citep[SMC;][]{Prevot1984} is adopted. Re-emitted grey-body dust is parameterized with a temperature of $\rm 100\,K$ and emissivity index of 1.6. 

The same modules and parametric space are used both for the X-ray catalogue and the reference galaxy catalogue. The AGN module is included in the SED fitting of the reference catalogue to identify sources with strong AGN component (see Section \ref{sec_sample_selection}). All free parameters used in the SED fitting process and their input values, are presented in Table \ref{table_cigale}. The grid has been optimized using only sources with spectroscopic redshift to avoid uncertainties and scatter introduced by the photoz measurements. 
 
X-CIGALE requires the intrinsic (X-ray absorption corrected) X-ray fluxes. We use the fluxes in the Hard energy band ($\rm 2-7\,keV$) and the correction factor, defined as the ratio between the observed and unabsorbed fluxes, provided in the \cite{Masini2020} catalogue, to estimate the intrinsic, hard X-ray flux and include it in the fitting process. The photon index, $\Gamma$, is fixed to 1.8. A maximal value of $|\Delta \alpha _{ox}|_{max} = 0.2$ is adopted for the dispersion of the $\rm \alpha _{ox} - L_{2500\AA}$ \citep{Risaliti2017} which corresponds to $\approx 2\sigma$ scatter in the $\rm \alpha _{ox} - L_{2500\AA}$ relation \citep{Just2007}.

We have verified that our results are robust and independent of e.g., adding more values for the slope, $\alpha$, of the Galactic dust emission in the \cite{Dale2014} model and the exact choice of the age of the stellar mass populations in the star forming history module. Most importantly, in our analysis, we use the SFR$_{norm}$ parameter to compare the SFR of X-ray sources with the SFR of non X-ray galaxies. Since SFR for both populations are calculated by X-CIGALE, using the same parameter space, we expect any biases and systematics, introduced by the grid, to be alleviated.

\begin{table*}
\caption{Number of X-ray AGN and sources in the reference galaxy catalogue.}
\centering
\setlength{\tabcolsep}{1.5mm}
\begin{tabular}{cccccc}
 \hline
&total &$\rm 0.5<z<1.0$ & $\rm 1.0<z<1.5$ & $\rm 1.5<z<2.0$ & $\rm 2.0<z<2.5$   \\
 & & $\rm log (M_*/M_\odot) > 10.73$ &$\rm log (M_*/M_\odot) > 11.19$ &$\rm log (M_*/M_\odot) > 11.52$ &$\rm log (M_*/M_\odot) > 11.43$\\
  \hline
  
X-ray catalogue &1,020 & 590& 298 & 90 & 42 \\
reference galaxy catalogue &18,248  & 14,993 & 2,956 & 262 &37   \\
  \hline
    \hline
X-ray catalogue* & 711 & 380& 247 & 84 &  \\

reference galaxy catalogue*  & 11,639& 9,171  & 2,229 & 239 &  \\ 
       \hline
\multicolumn{6}{l}{\footnotesize{*Excluding quiescent galaxies (see text for more details)}}
\label{table_data}
\end{tabular}
\end{table*}

\begin{figure}
\centering
  \includegraphics[width=1.\linewidth, height=6cm]{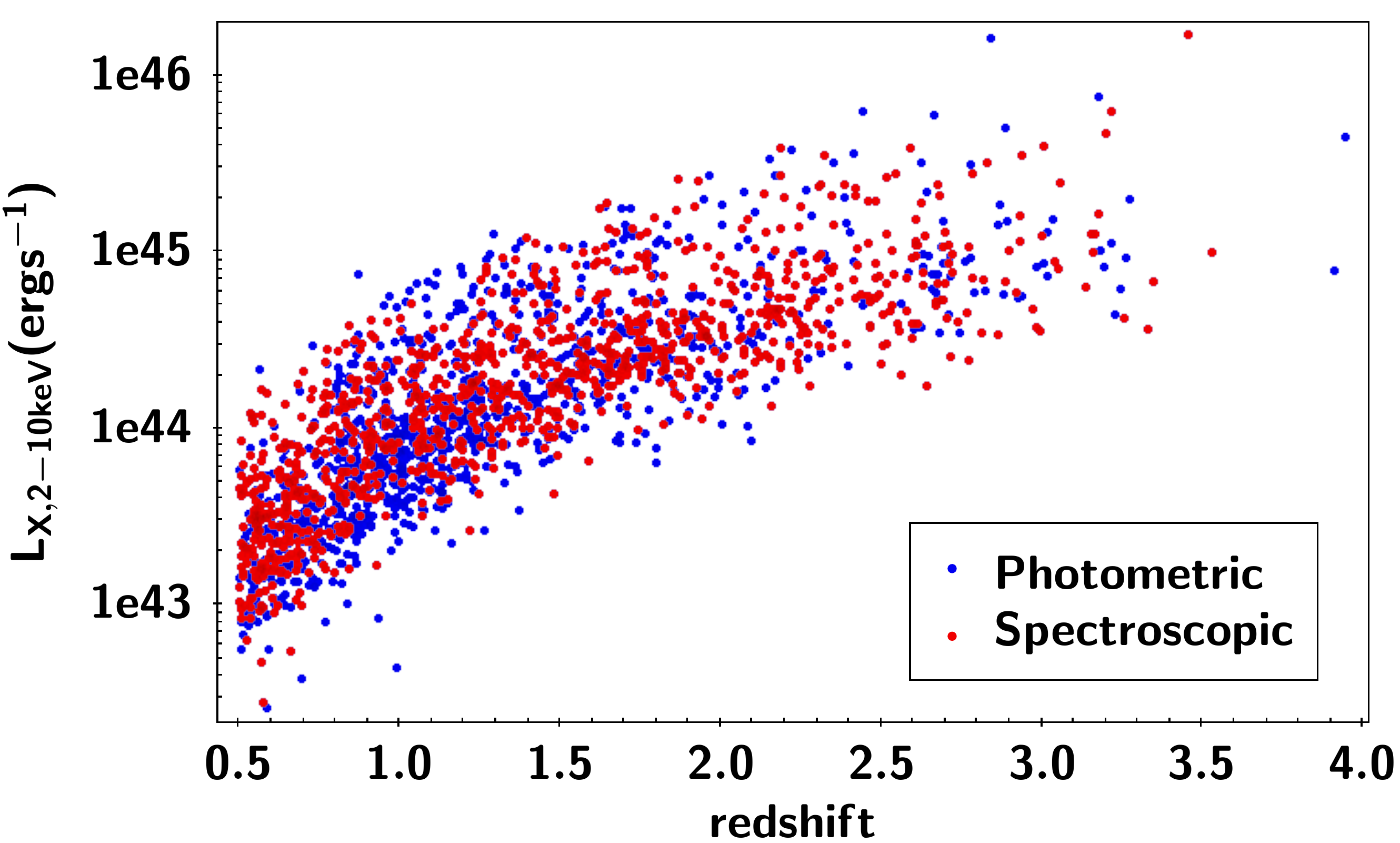}
  \caption{The X-ray hard, instrinsic luminosity as a function of redshift, for the 1989 X-ray AGN in our sample. There are 1020 sources with spectroscopic redshifts (red circles) and 969 with photometric redshifts (blue circles).}
  \label{fig_redz_lx}
\end{figure}

\begin{figure}
\centering
  \includegraphics[width=1.\linewidth, height=6cm]{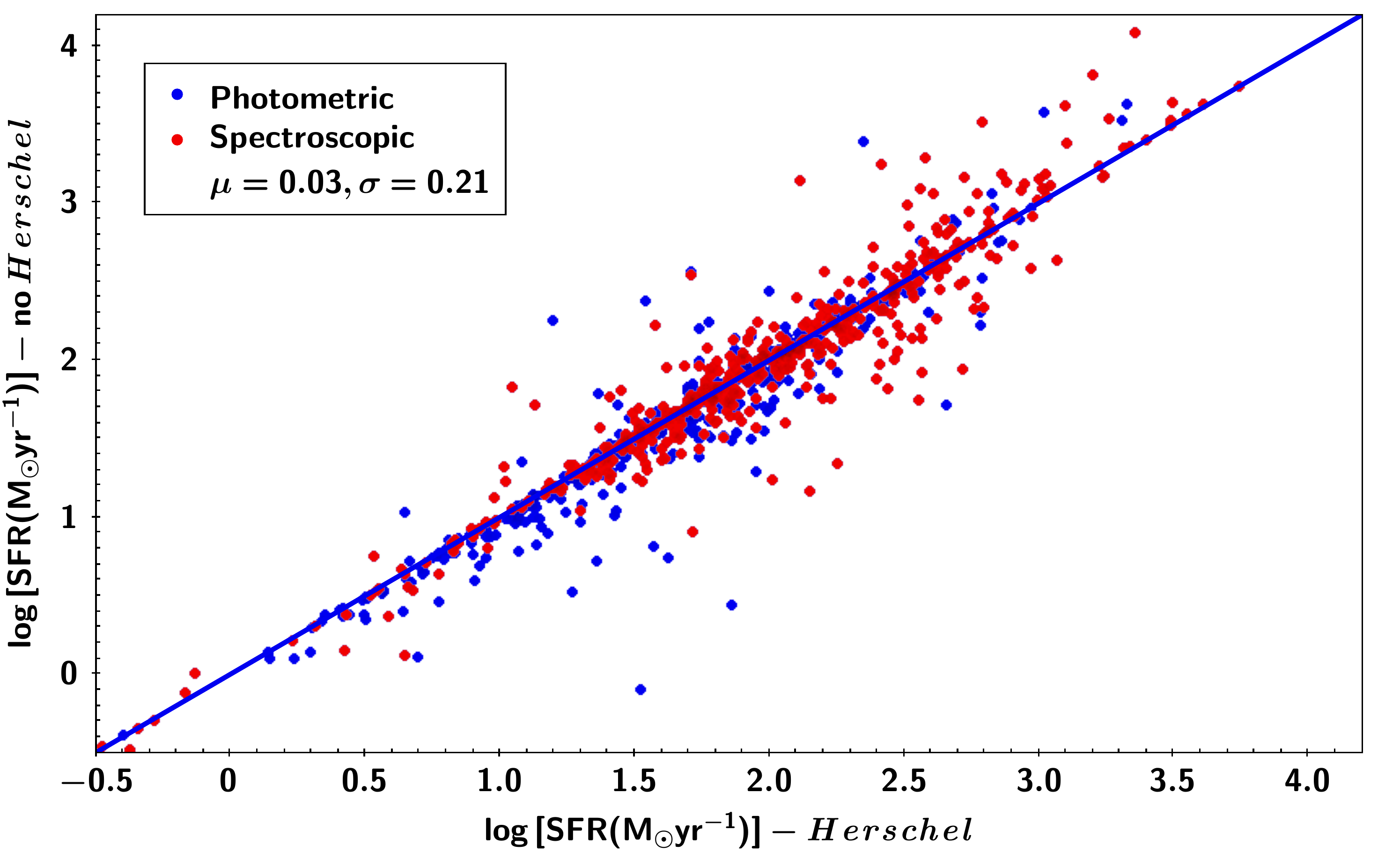}
  \caption{Comparison of SFR measurements with and without {\it{Herschel}} photometry, for 935 sources that have far-IR coverage and satisfy our selection criteria (see text for more details). X-ray AGN with spectroscopic redshifts (490) are shown in red while sources with photometric redshift estimations (445) are presented in blue. Blue solid line presents the 1:1 relation. Results show an excellent agreement between the two measurements. The mean offset and the standard deviation, of the SFR difference (${\it{Herschel}}-\rm{no}\,{\it{Herschel}}$) are shown in the legend.}
  \label{fig_herschel}
\end{figure}

\begin{figure*}
\centering
\begin{subfigure}{.505\textwidth}
  \centering
  \includegraphics[width=.92\linewidth, height=6.2cm]{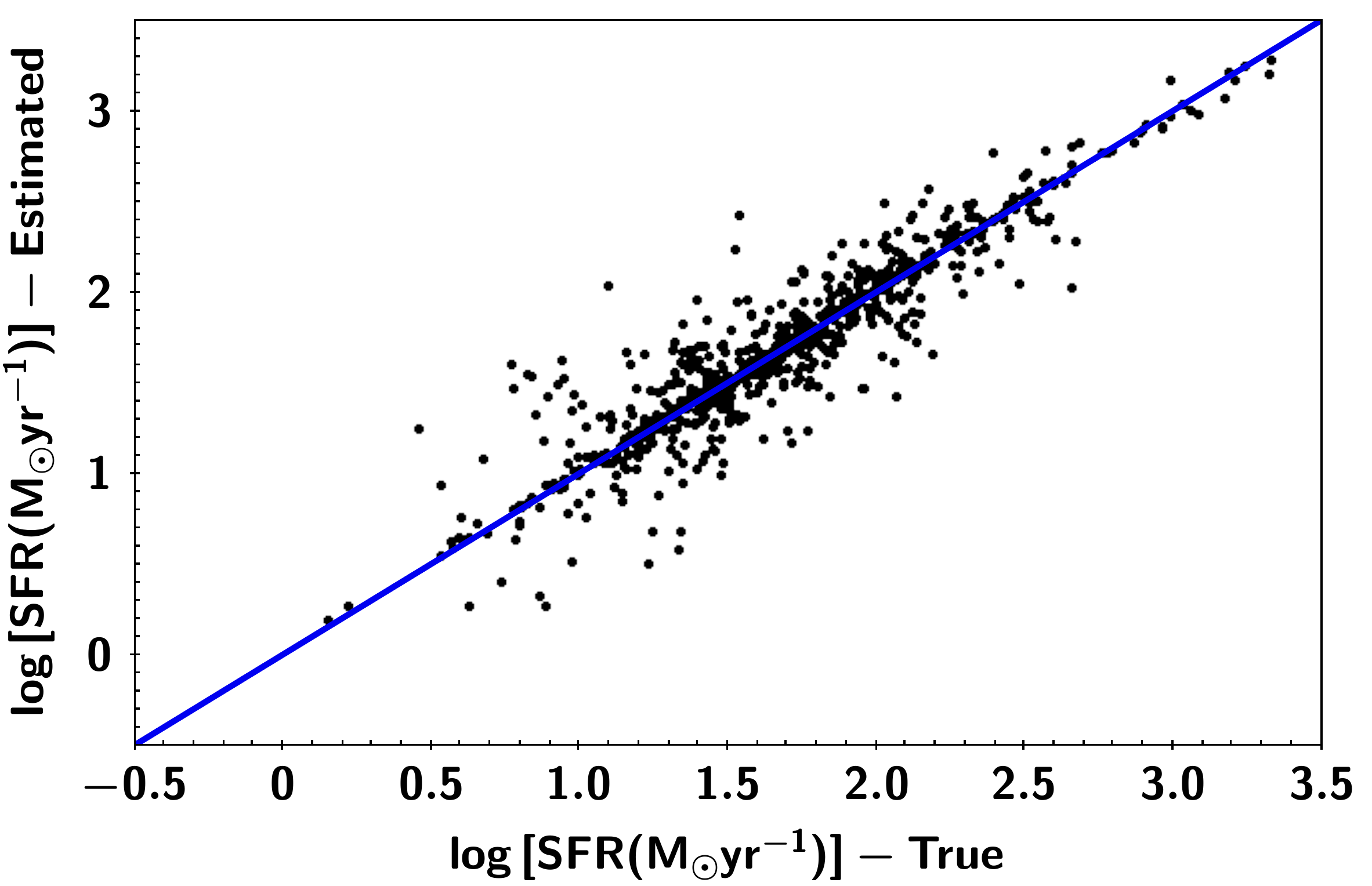}
%  \caption{}
  \label{}
\end{subfigure}%
\begin{subfigure}{.52\textwidth}
  \centering
  \includegraphics[width=.92\linewidth, height=6.2cm]{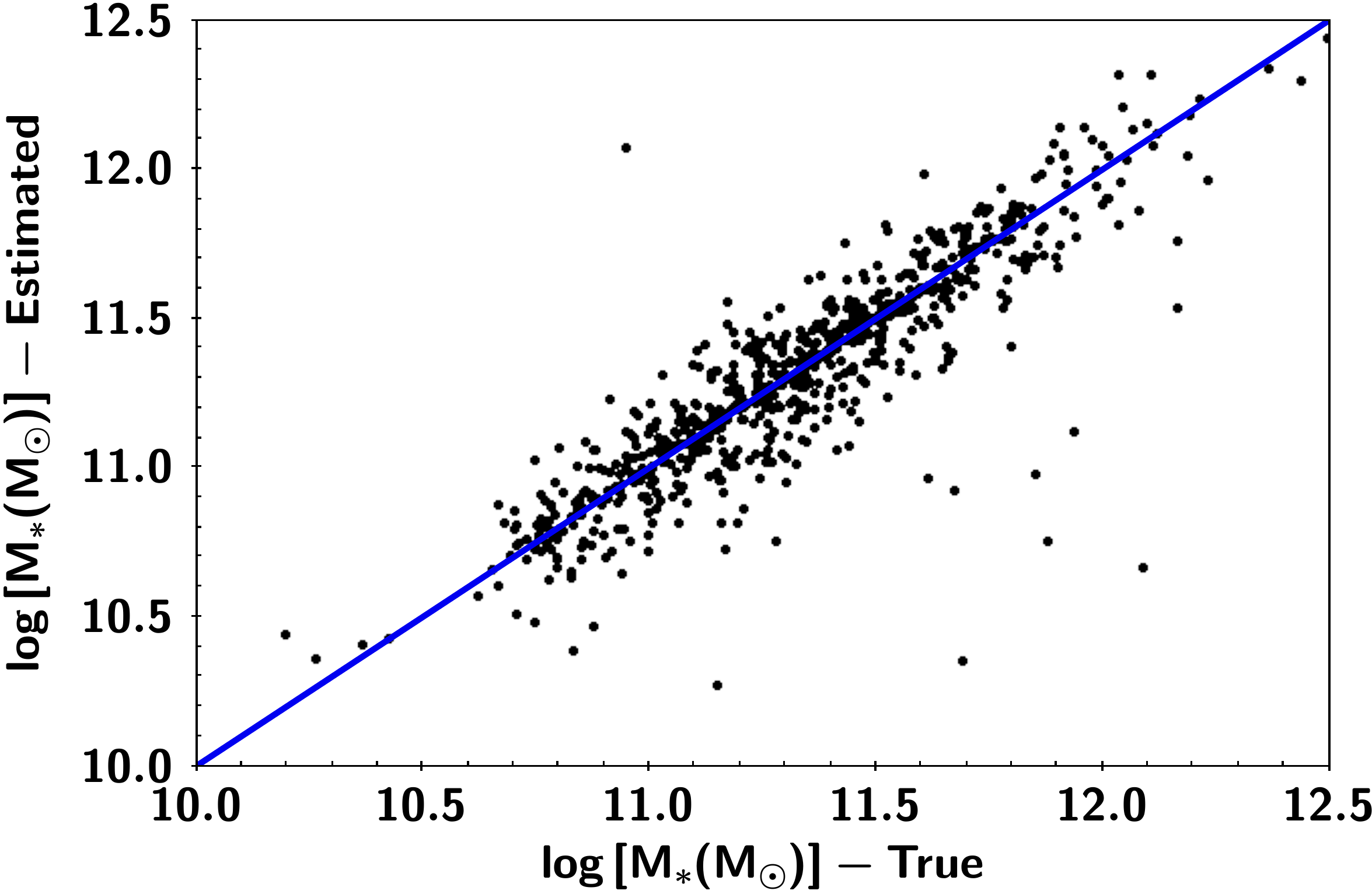}
%  \caption{}
  \label{}
\end{subfigure}
\caption{Comparison of the SFR and M$_*$ measurements (left and right panel, respectively) for the mock sources (estimated) with the true values (data), using the X-ray sample. X-CIGALE accurately recovers the true SFR and M$_*$ values of the mock sources.}
\label{fig_data_mock}
\end{figure*}

\begin{figure*}
\centering
\begin{subfigure}{.505\textwidth}
  \centering
  \includegraphics[width=.92\linewidth, height=6.2cm]{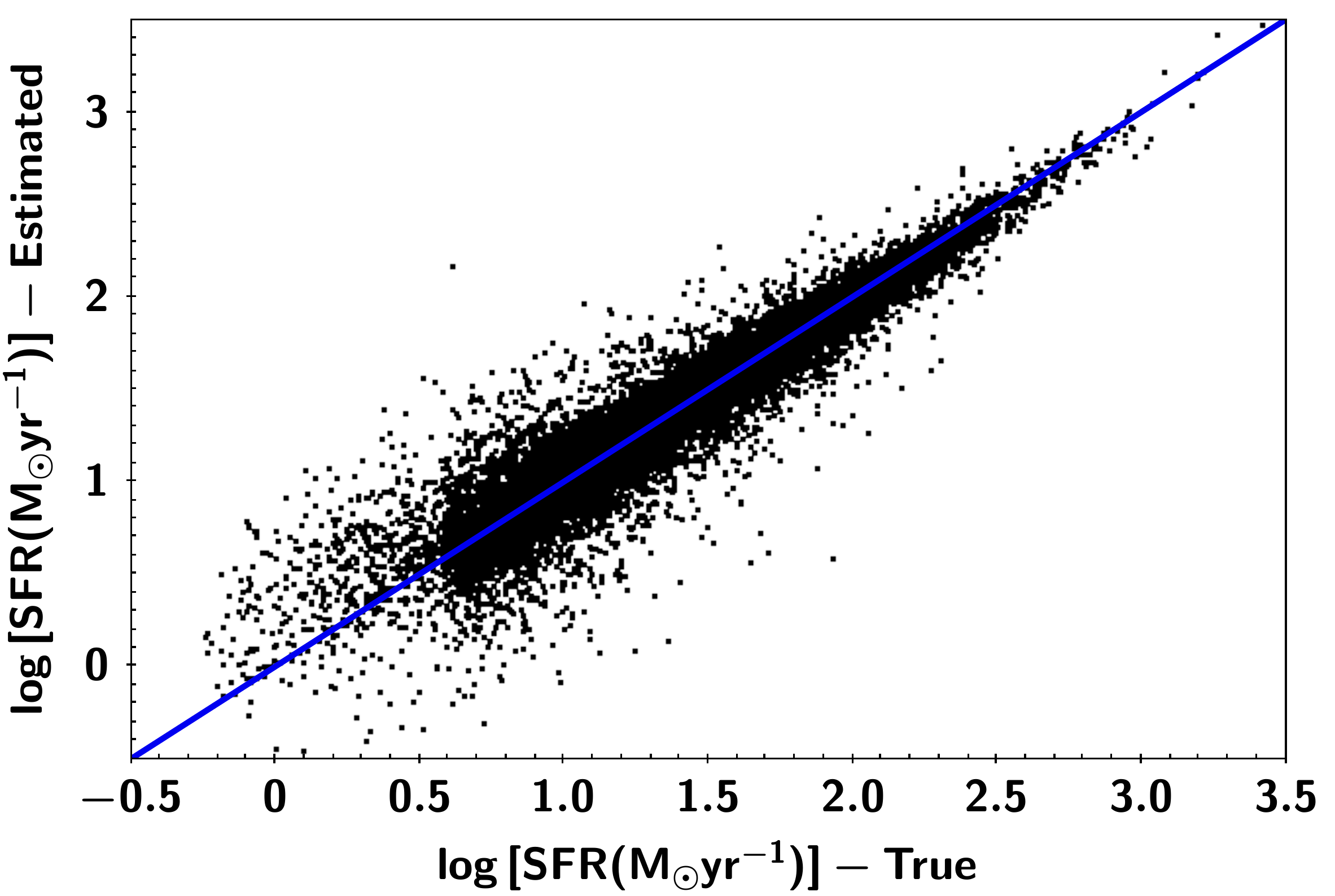}
%  \caption{}
  \label{}
\end{subfigure}%
\begin{subfigure}{.52\textwidth}
  \centering
  \includegraphics[width=.92\linewidth, height=6.2cm]{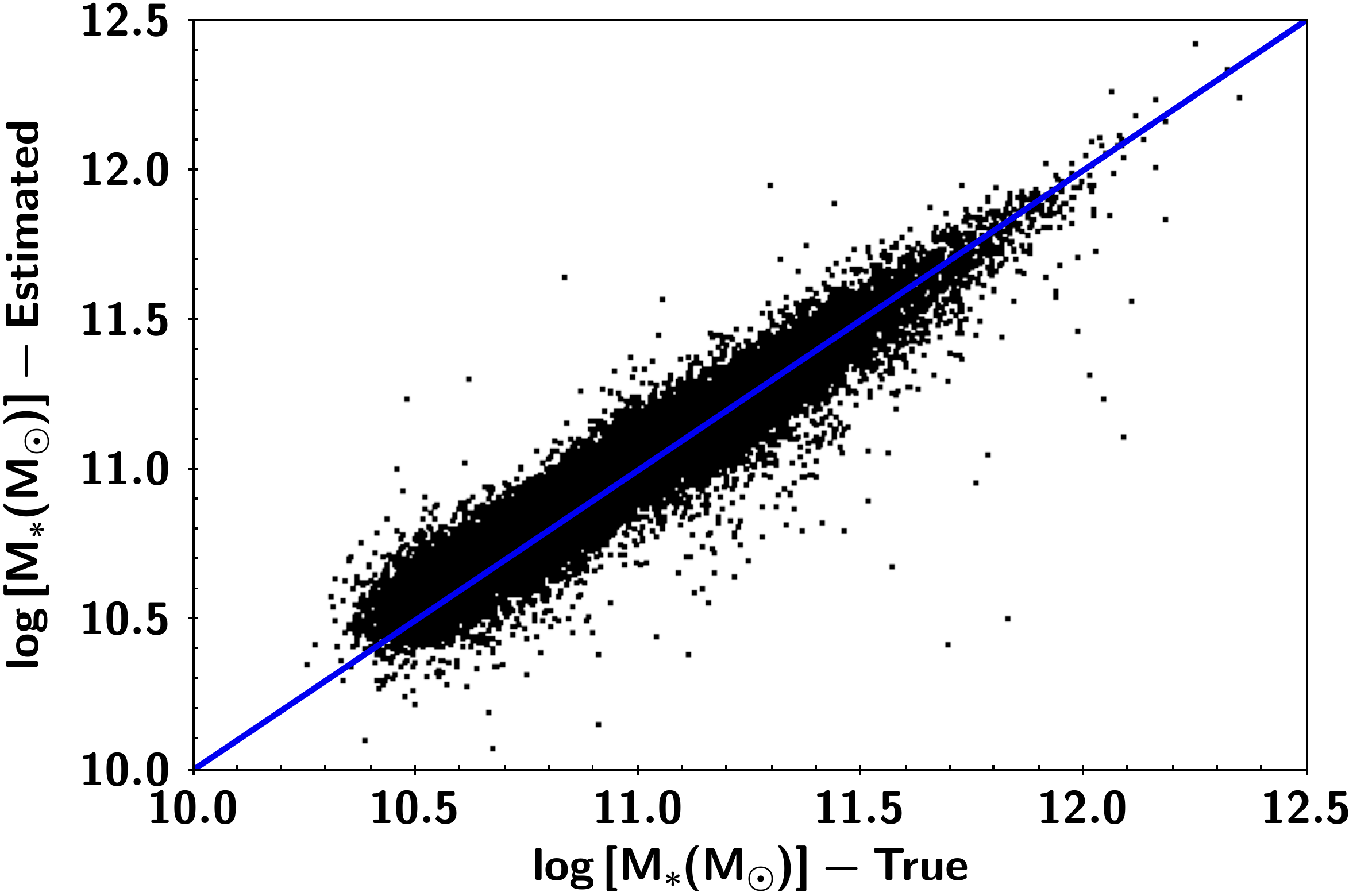}
%  \caption{}
  \label{}
\end{subfigure}
\caption{Comparison of the SFR and M$_*$ measurements (left and right panel, respectively) for the mock sources (estimated) with the true values (data), using the reference galaxy sample. X-CIGALE accurately recovers the true SFR and M$_*$ values of the mock sources.}
\label{fig_data_mock_galaxy}
\end{figure*}

\subsection{Exclusion of sources with bad fits}
\label{sec_bad_fits}

X-CIGALE provides two estimates for each of the measured galaxy properties. One that is evaluated from the best-fit model and one that weighs all models allowed by the parametric grid, with the best-fit model having the heaviest weight \citep{Boquien2019}. This weight is based on the likelihood, exp ($-\chi^2/2)$, associated with each model. A large difference between these two values indicates that the probability density function (PDF) is asymmetric and a simple (e.g. Gaussian) model for the errors is not valid. To exclude from our datasets such cases that result in unreliable SFR and stellar mass estimations, we keep both from the X-ray and the reference galaxy catalogues, only systems with $\rm \frac{1}{5}\leq \frac{SFR_{best}}{SFR_{bayes}} \leq 5$ and $\rm \frac{1}{5}\leq \frac{M_{*, best}}{M_{*, bayes}} \leq 5$, where SFR$\rm _{best}$, M$\rm _{*, best}$ are the best fit values of SFR and M$_*$, respectively and SFR$\rm _{bayes}$ and M$\rm _{*, bayes}$ are the Bayesian values, estimated by X-CIGALE. These reduce the number of available X-ray AGN to 1989 (from 2338) and the number of sources in the reference galaxy catalogue to 52,531 (from 56,627). Varying the boundaries of the criterion within a range of $0.1-0.33$ for the lower limit and $3-10$ for the upper limit changes the size of our samples by up to $\pm0.05\%$ and therefore does not affect the results of our analysis. Fig. \ref{fig_redz_lx} presents the X-ray luminosity as a function of redshift. Specz sources (red circles) have, on average, slightly higher X-ray luminosities (median $\rm L_{X,2-10keV}\sim 10^{44.5}\,ergs^{-1}$) compared to photoz sources (median $\rm L_{X,2-10keV} \sim 10^{44}\,ergs^{-1}$) while both populations lie at similar redshifts.

\subsection{Mass completeness}
\label{sec_mass_completeness}

In our analysis, we split the sources into four redshift bins, from 0.5 to 2.5, with a bin size of 0.5. This allows us to study separately possible dependence of the SFR of X-ray AGN with redshift and X-ray luminosity. To avoid any biases introduced by mass incompleteness of our samples, we estimate the mass completeness of the X-ray and reference galaxy catalogues at each redshift bin. For that purpose, we use the method described in \cite{Pozzetti2010}. We estimate the limiting stellar mass (M$\rm _{lim}$) of each galaxy at each redshift, which is the mass the galaxy would have if its apparent magnitude was equal to the limiting magnitude of the survey for a specific photometric band. M$\rm _{lim}$ is calculated by the following expression:

\begin{equation}
\rm log M_{lim} = log M_*+0.4(m-m_{lim}),
\end{equation}
where m is the AB magnitude of the source and m$_{\rm lim}$ is the AB magnitude limit of the survey. The result is a distribution of $\rm log M_{lim}$ that reflects the distribution of stellar mass-to-light ratio ($\rm M/L$), at each redshift. To obtain a representative mass limit of our dataset, we use the $\rm log M_{lim}$ of the $20\%$ faintest galaxies at four redshift bins. This effectively removes galaxies with the highest $\rm M/L$ that do not significantly contribute close to the magnitude limit of the survey. Then, the minimum stellar mass at each redshift interval for which our sample is complete is the 95th percentile of $\rm log M_{lim}$, of the $20\%$ faintest galaxies in each redshift bin. K$_s$ is the limiting band of our sample and has been chosen to define the mass completeness of our dataset ($20\%$ of the sources in the parent HELP catalogue have detection in this band). Moreover, the K$_s$ band is shallow enough that essentially all galaxies with a detection in this band are also detected in bluer bands.

Following the aforementioned procedure, and using the K$_s$ band\footnote{\url{http://hedam.lam.fr/HELP/dataproducts/dmu0/dmu0_IBIS/readme.md/}} with $ \rm m_{lim}= 21.35\,mag$, we find that the stellar mass completeness for our galaxy sample is defined at $\rm log\,[M_{*,95\%lim}(M_\odot)]= 10.73, 11.19, 11.52, 11.43$ at $\rm 0.5<z<1.0$, $\rm 1.0<z<1.5$, $\rm 1.5<z<2.0$ and $\rm 2.0<z<2.5$, respectively. Using a different photometric band that traces the stellar mass at high redshift, for instance the IRAC1 or IRAC2 bands, changes the above mass completeness limits by $<0.15\,\rm dex$. Finally, we note that these stellar mass limits are quite high. Equation (9) of \cite{Schreiber2015} that is used to estimate SFR$_{norm}$ in Section \ref{sec_using_schreiber}, has been calibrated for sources up to $\rm log\,[M_*(M_\odot)]=11.5$.

\subsection{Final samples}
\label{sec_sample_selection}

In Section \ref{sec_using_schreiber}, we estimate SFR$_{norm}$ applying the Schreiber et al. equation. For that, we use the X-ray sample defined in Section \ref{sec_bad_fits}, i.e. the 1989 X-ray selected AGN. This allows us to perform a fair comparison with previous studies that used the same equation and did not apply any mass completeness criteria on their X-ray sources. This is also the X-ray sample used in Section \ref{sec_host_properties}, where we study the host galaxy properties of X-ray absorbed and unabsorbed AGN. However, in Section \ref{sfrnorm_using_control_sample}, where we examine the relation of the AGN with the SFR of the host galaxy, we apply the mass completeness limits we estimated in the previous Section, on both the X-ray and reference galaxy catalogues (in addition to the cuts described below). 

As already noted, most previous X-ray studies used the \cite{Schreiber2015} equation to calculate the SFR of star forming MS galaxies and explore the SFR of X-ray AGN relative to main sequence star forming galaxies. In Schreiber et al. they do not explicitly exclude AGN from their star forming galaxy sample. Since there is no AGN template in their fitting analysis, most AGN systems will be badly fitted. Thus, they reject sources for which the SED fits have large $\chi ^2$ values ($\chi ^2 >10$). In our analysis, we include an AGN template when we fit the X-ray and the reference galaxy catalogues (see Section \ref{sec_cigale}). This enables us to identify systems with an AGN component and quantify it. We use the $\rm frac_{AGN}$ parameter to exclude such systems from the reference galaxy catalogue. Specifically, we reject sources with $\rm frac_{AGN}>0.2$. These systems account for $\approx 14\%$ of the total reference galaxy sample. The percentage ranges from $\sim 14\%$ at $\rm z<1.5$, up to $\sim 50\%$ at higher reshifts. This is qualitatively and quantitatively consistent with observational studies that found higher AGN duty cycles at earlier epochs and for more massive galaxies \citep[e.g.][]{Genzel2014, Georgakakis2017b}. 

Application of the mass completeness criteria in both the X-ray and the reference galaxy catalogues and the exclusion of systems with a (strong) AGN component from the reference sample, results in 1020 X-ray AGN and 18,248 sources in the reference galaxy catalogue. The number of available sources in the two datasets are presented in Table \ref{table_data}.

\subsection{Reliability of galaxy properties measurements}
\label{sec_reliability}

From the 1989 X-ray AGN, 935 ($\sim 47\%$) have been detected by {\it{Herschel}}. In the analysis, both 100 and 160 $\mu$m PACS bands and SPIRE 500$\mu$m band are not considered, given their lower sensitivity for this field \citep{Oliver2012}. For these sources, we perform SED analysis with and without {\it{Herschel}} bands, using the same parameter space in the fitting process. The comparison of the SFR measurements is presented in Fig. \ref{fig_herschel}. To examine whether photoz introduce (additional) scatter to the comparison of the SFR calculations, we plot sources with specz in red and those with photoz in blue. We find a very good agreement of the SFR values between the two measurements ($\rm SFR_{no\,{\it{Herschel}}}=(1.02\pm0.01)\,SFR_{{\it{Herschel}}}-0.07\pm0.01$). We also note that sources with photoz do not present larger dispersion compared to spectroscopic sources. This result shows that the SFR calculations of those sources without far-IR photometry in our dataset, are reliable. 

We, then, examine the accuracy and reliability of the SFR and M$_*$ measurement of X-CIGALE. We quantify the accuracy using the $sigma$ parameter ($\rm {\it{sigma}}=\frac{value}{error}$). For the X-ray catalogue, the average $sigma$ is  $\sim 3.7$ and $\sim 4.5$, for the SFR and M$_*$ measurements, respectively. For the reference galaxy catalogue, $sigma$ is  $\sim 4.2$ and $\sim 5.8$, for the SFR and M$_*$ measurements, respectively. Stellar mass and SFR calculations of sources in the reference galaxy catalogue, are more robust than those for X-ray AGN. This is due to the AGN emission that can outshine the optical emission of the host galaxy, thus rendering measurements less accurate for AGN, in particular for unobscured/type 1 AGN.
 
X-CIGALE offers the option to create and analyse mock catalogues based on the best fit model of each source of the dataset. When this option is chosen, the code uses the best fit of each source and creates a mock sample. Each best flux is modified by injecting noise extracted from a Gaussian distribution with the same standard deviation as the observed flux. Then, the mock data are analysed in the same way as the observed data. The precision of each estimated parameter can be tested by comparing the input and output values (ground truth vs. estimated value). We use these mock catalogues to examine the reliability of SFR and M$_*$ results. In Fig. \ref{fig_data_mock} (left panel), we plot the comparison of the Bayesian values of SFR obtained from the fit of the mock catalogue with the true SFR values from the best fit of the X-ray sample. The right panel shows the results of the stellar mass measurements. Fig. \ref{fig_data_mock_galaxy} shows the comparison for the reference galaxy sample. The scatter in the stellar mass measurements is larger for X-ray AGN compared to the reference sample. As, mentioned earlier this is due to the AGN component. In both cases, X-CIGALE successfully recovers the true SFR and M$_*$ values of the mock sources.

In the analysis presented in Section \ref{sec_lx_sfr}, we take into account the uncertainties of the SFR and M$_*$, estimated by X-CIGALE. Specifically, in the analysis based on SFR$_{norm}$, each source is weighted by the $sigma_{SFR} \times sigma_{M_*}$, while for the analysis based on $\lambda_{s,BHAR}$, each source is weighted using the value of $sigma_{M_*}$.

\begin{figure}
\centering
\begin{subfigure}{.5\textwidth}
  \centering
  \includegraphics[width=1.\linewidth, height=5.cm]{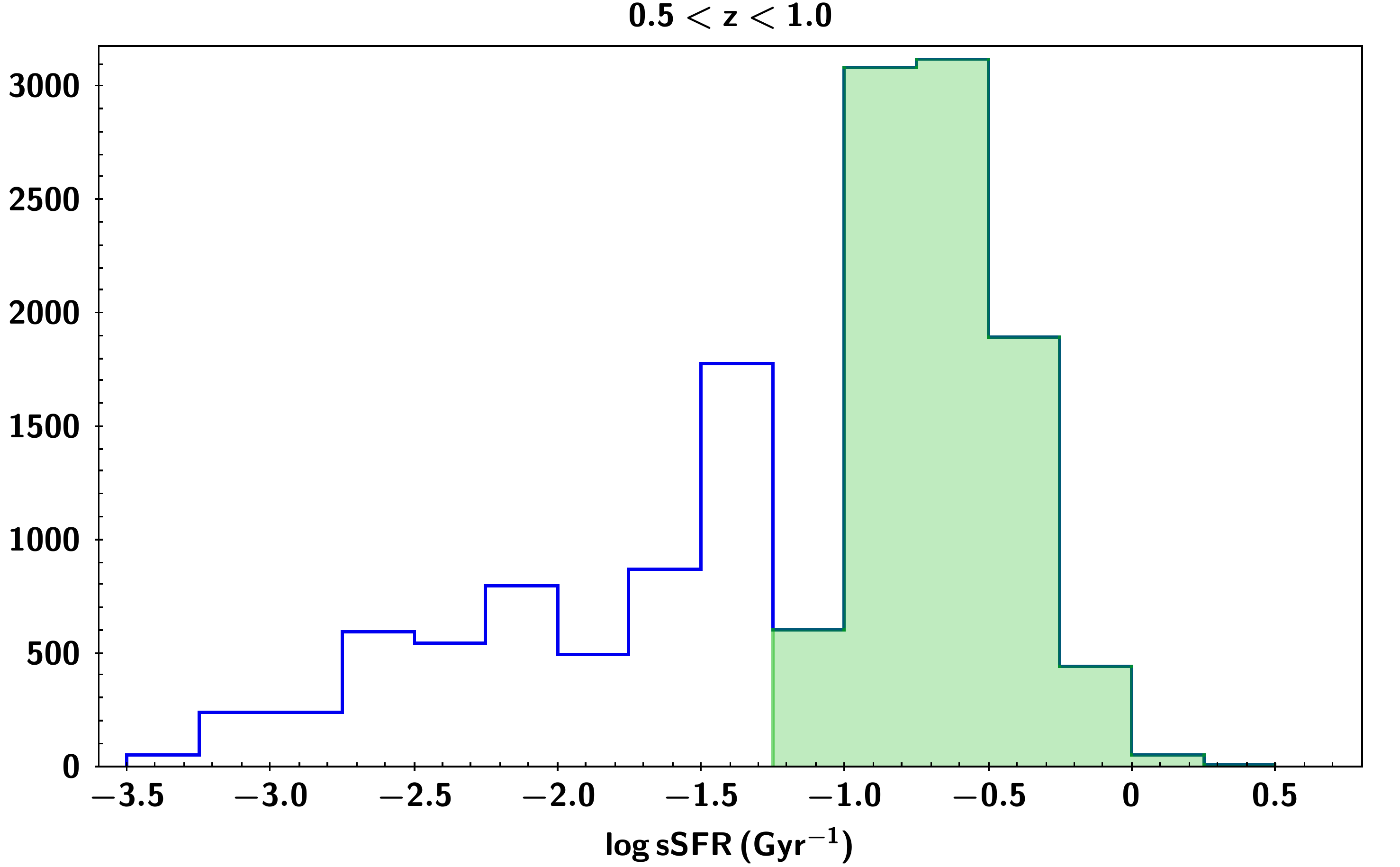}
%  \caption{}
  \label{}
\end{subfigure}
\begin{subfigure}{.5\textwidth}
  \centering
  \includegraphics[width=1.\linewidth, height=5.cm]{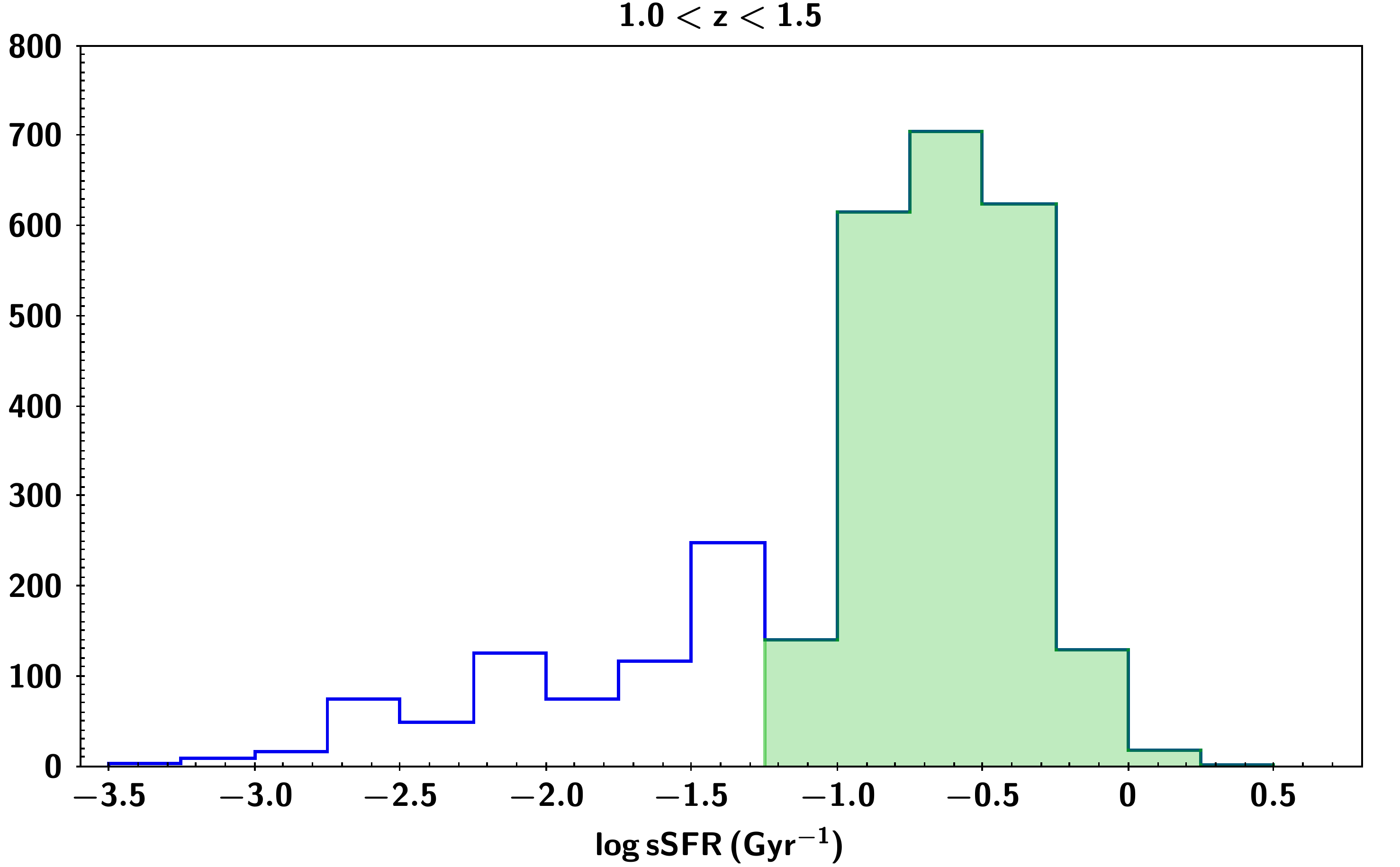}
%  \caption{}
  \label{}
\end{subfigure}
\begin{subfigure}{.5\textwidth}
  \centering
  \includegraphics[width=1.\linewidth, height=5.cm]{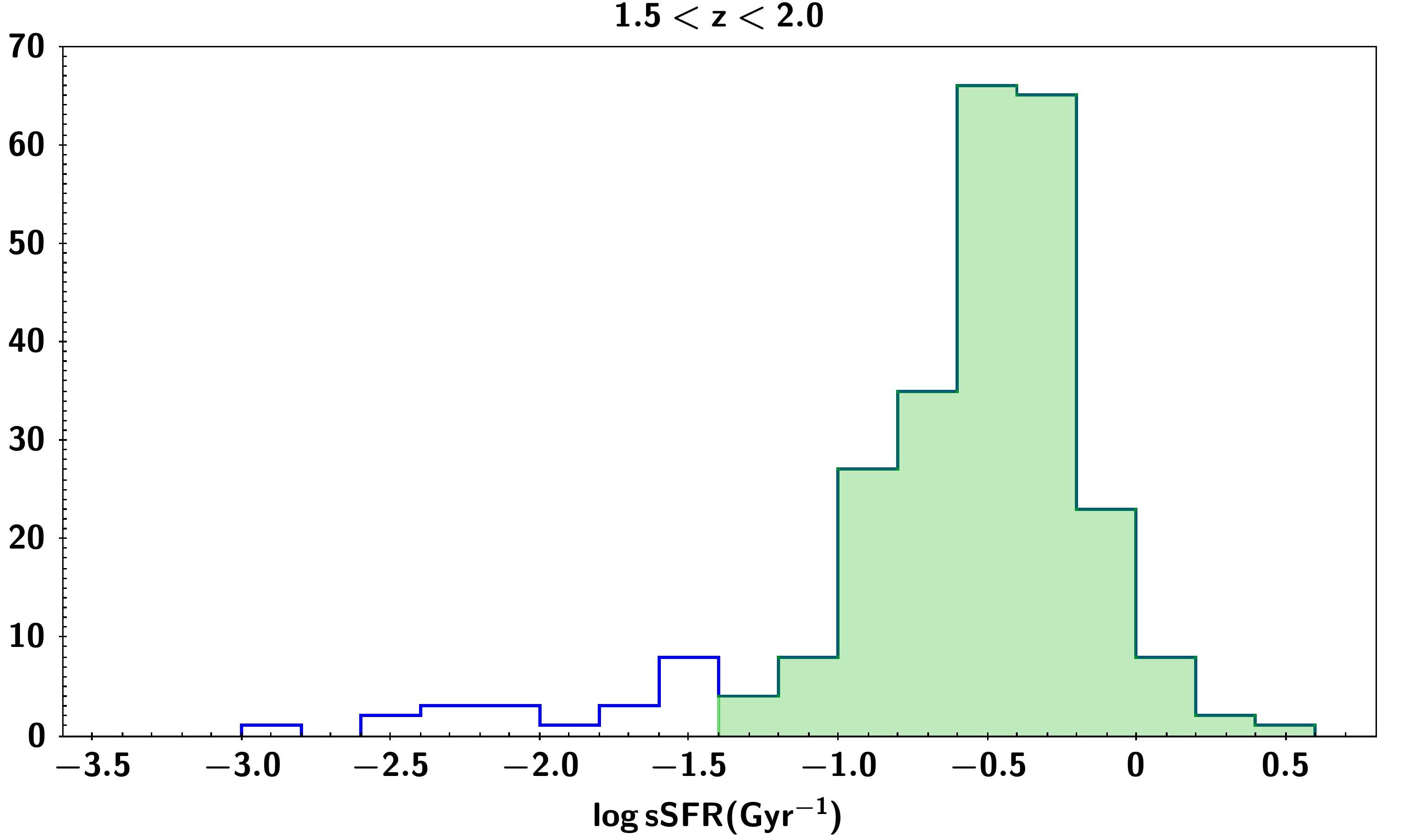}
%  \caption{}
  \label{}
\end{subfigure}
\begin{subfigure}{.5\textwidth}
  \centering
  \includegraphics[width=1.\linewidth, height=5.cm]{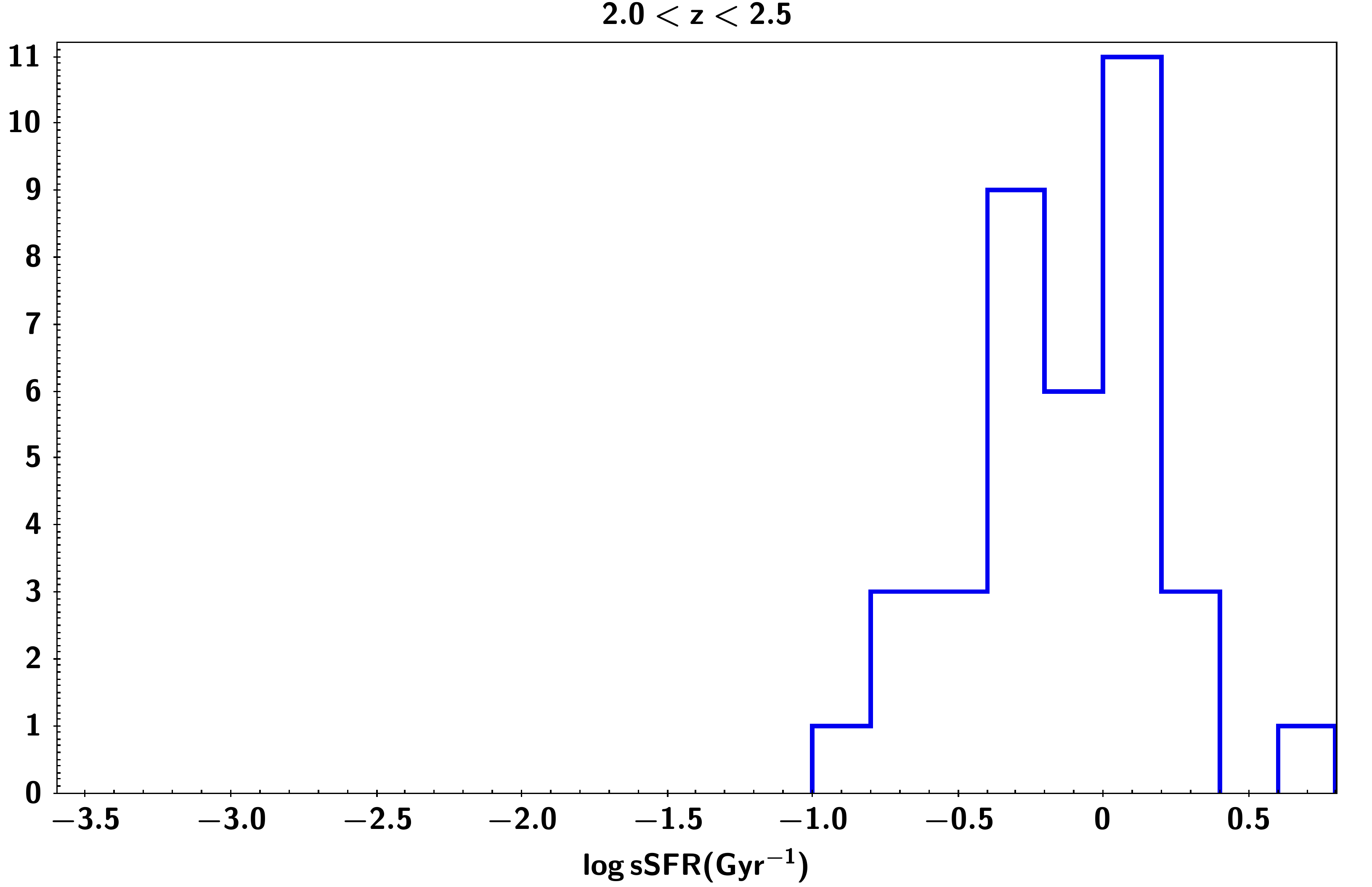}
%  \caption{}
  \label{}
\end{subfigure}
\caption{sSFR distribution in four redshift intervals. Blue lines present the full distributions. Green areas are the sSFR distribution after applying the sSFR cut, which is defined based on the location of the second, lowest peak of each distribution. At the highest redshift bin ($\rm 2.0<z<2.5$), the sSFR distribution does not have a Gaussian like shape, most likely due to the low number of sources (see Table \ref{table_data}) and the sSFR cut cannot defined as described above.}
\label{fig_ssfr}
\end{figure}

\begin{figure*}
\centering
\begin{subfigure}{.505\textwidth}
  \centering
  \includegraphics[width=.92\linewidth, height=6.2cm]{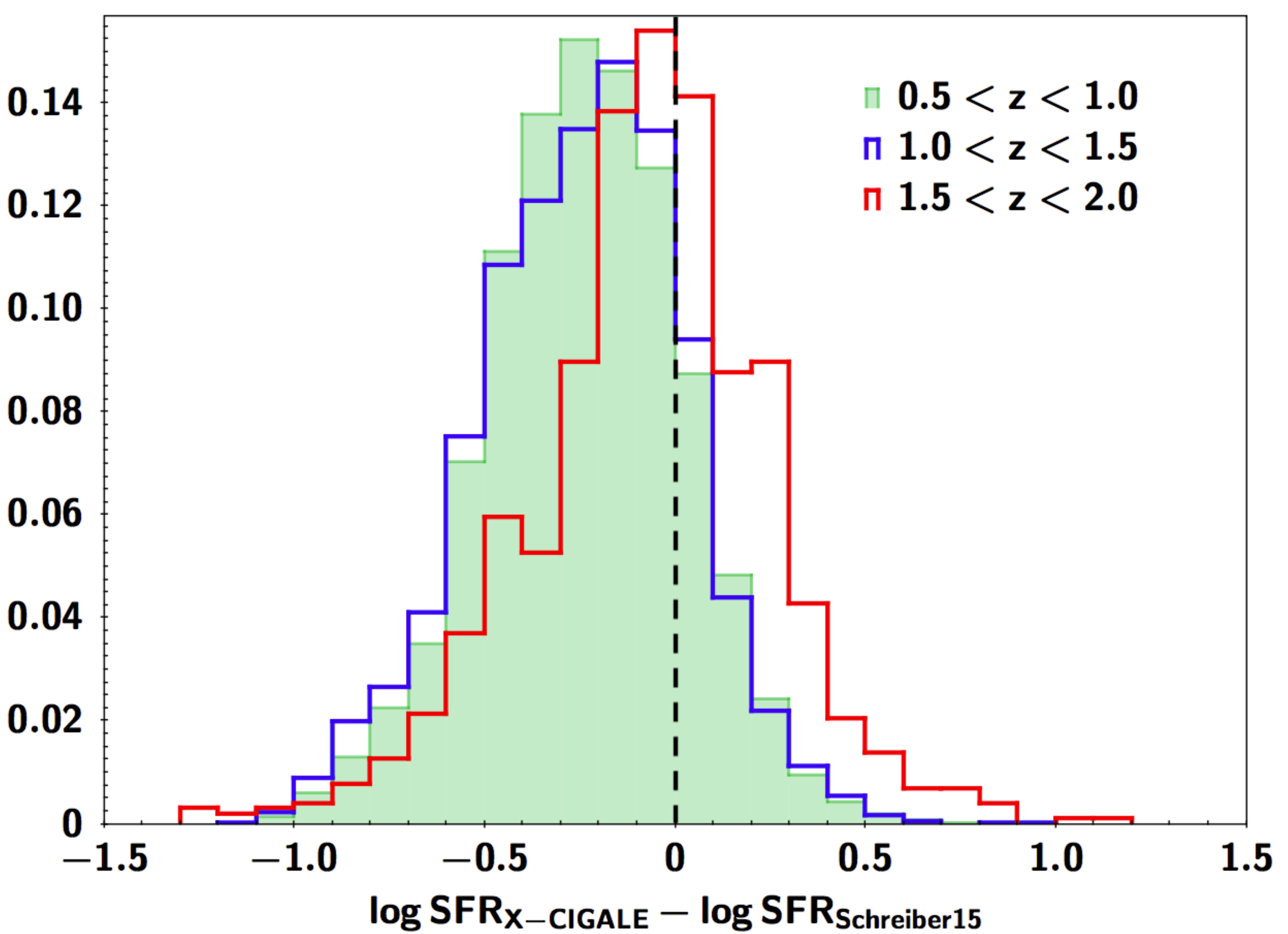}
%  \caption{}
  \label{}
\end{subfigure}%
\begin{subfigure}{.52\textwidth}
  \centering
  \includegraphics[width=.92\linewidth, height=6.2cm]{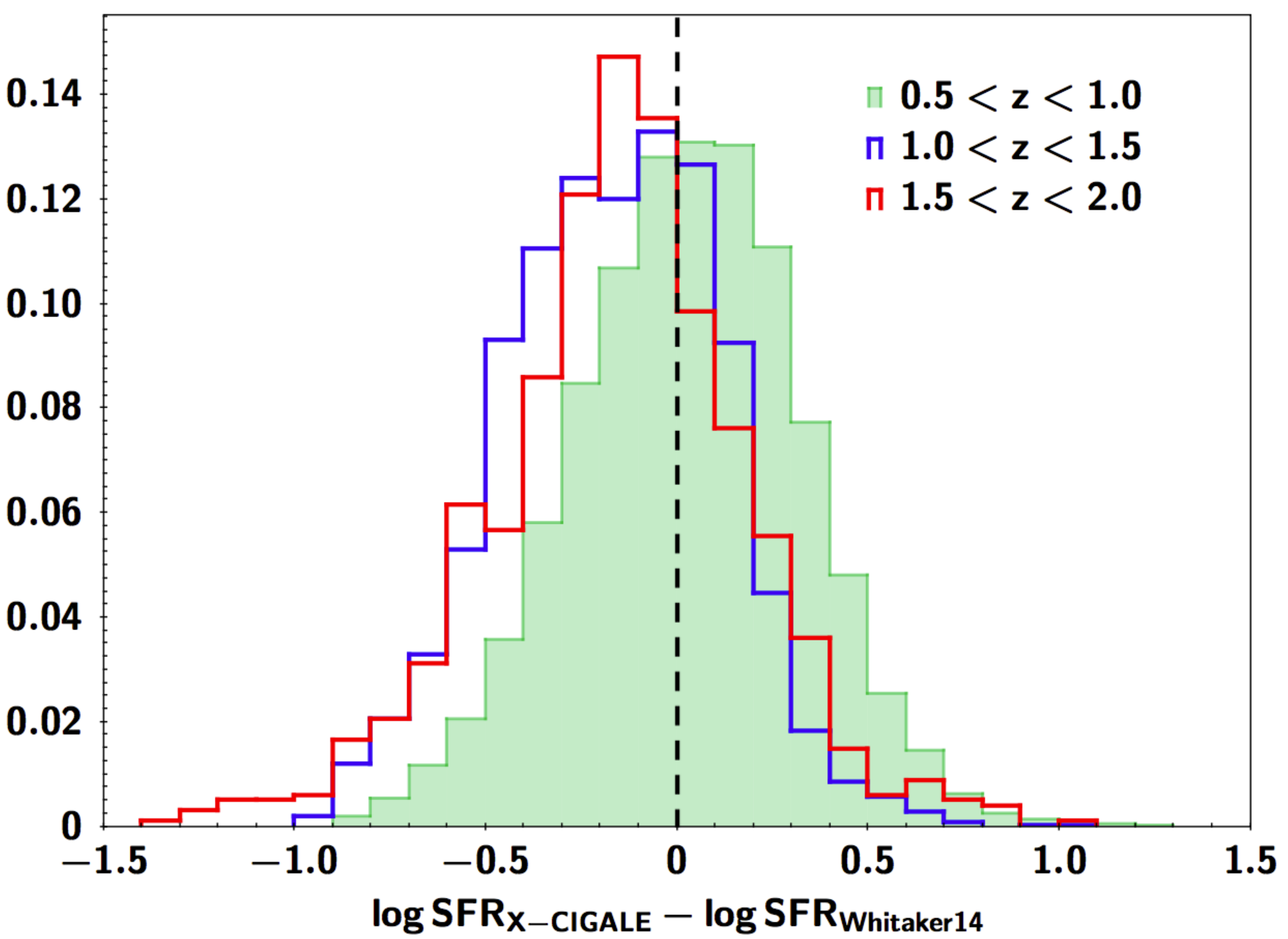}
%  \caption{}
  \label{}
\end{subfigure}
\caption{Left: Difference of the SFR calculations of X-CIGALE for star forming galaxies defined from our galaxy sample (see text for more details) and the SFR measurements using the expression of Schreiber et al. for star forming MS galaxies. At $\rm z<1.5$ SFR values from X-CIGALE are $\sim 0.25$\,dex lower compared to those using Schreiber et al., whereas at higher redshift the two estimates are consistent. Right: The difference of the SFR calculations of X-CIGALE for star forming galaxies defined from our galaxy sample (see text for more details) and the SFR measurements using the expression of \cite{Whitaker2014} for star forming MS galaxies. At $\rm z<1.0$, SFR values from X-CIGALE are similar (distribution peaks at zero) to those using the Whitaker et al. formula, whereas at $\rm 1<z<2$, SFR$_{\rm X-CIGALE}$ are lower by $\sim 0.15$\,dex (see also Table \ref{table_delta_sfr}).}
\label{fig_cigale_vs_schreiber}
\end{figure*}

\section{Definition of Main Sequence: systematics and selection effects}
\label{sec_systematics}

The goal of this work is to examine whether there is a correlation between the AGN power and the SFR of the host galaxy. The SFR of X-ray AGN evolves with stellar mass and redshift in a way that is qualitatively similar to the SFR evolution of star forming galaxies \citep[][]{Masoura2018}. To account for this evolution, previous studies estimate the SFR$_{norm}$ parameter, that is defined as the ratio of the SFR of X-ray AGN to the SFR of star forming MS galaxies, $\rm SFR_{{\it{norm}}}=\frac{SFR_{AGN}}{SFR_{MS}}$, \citep{Mullaney2015, Masoura2018, Bernhard2019, Aird2019, Grimmett2020, Masoura2021}. For that purpose, nearly all of them, use the analytical expression (9) of \cite{Schreiber2015} to parametrize the SFR of main-sequence galaxies. However, this method hints at a number of systematics and uncertainties. X-ray and non X-ray galaxy samples are defined differently  and different methods are applied to estimate galaxy properties and specifically SFR and stellar masses which are used to define the Main Sequence.

To acquire better control of these systematics, in our analysis on both the X-ray and the galaxy reference sample we have applied the same photometric criteria. Furthermore, SFR (and M$_*$) of both datasets have been measured using the same SED fitting method, by applying the same parametric grid. We also define our own MS for star forming galaxies, using our data. Towards this end, we identify and exclude quiescent galaxies from our datasets. Our goal is not to make a strict definition of star forming MS galaxies, but mainly to exclude in a uniform manner (most) of the quiescent systems from both the X-ray and the reference galaxy catalogues, by applying the same selection criteria in the two samples. For that purpose, we estimate the sSFR, defined as the $\frac{SFR}{M_*}$ of each source that is above the mass completeness limit at the corresponding redshift (Section \ref{sec_mass_completeness}). We use only the galaxy reference sample to define sSFR criteria at different redshift bins and exclude quiescent galaxies, due to its significantly larger size compared to the X-ray sample. These sSFR criteria are then used to exclude quiescent systems from the X-ray sample, too. 

The sSFR distributions are presented in Fig. \ref{fig_ssfr}. Up to $\rm z<2$ a tail and a second, smaller peak is present. The quiescent population of the galaxy sample, is defined by applying a sSFR cut at the location of this second peak of each distribution. At the highest redshift bin, the sSFR distribution does not appear Gaussian, probably due to the small number of sources, and the definition of the sSFR cut is not obvious. In the remaining of our analysis we will not include sources from that redshift bin. The above sSFR cuts exclude $\sim 35\%$ of the sources, as quiescent galaxies (Table \ref{table_data}). Most of them lie at $\rm z<1.5$ ($\sim 40\%$ at $\rm 0.5<z<1.0$ and $\sim 25\%$ at $\rm 1.0<z<1.5$) while $\sim 10\%$ of the galaxies in the reference catalogue are classified as quiescent systems at $\rm z>1.5$. This is consistent with studies that examine the evolution of quiescent and star forming galaxies \citep[e.g.][]{Bezanson2012}.

%and find evolution of quiescent galaxies with redshift, presumably due to star forming galaxies shutting off their star formation \citep[e.g.][]{Bezanson2012}.

Fig. \ref{fig_cigale_vs_schreiber} illustrates the systematic difference between the SFR calculations of X-CIGALE for star forming galaxies defined from our galaxy sample, as described above, and the SFR measuremments using the expression of Schreiber et al. for star forming MS galaxies. At $\rm z<1.5$ SFR values from X-CIGALE are $\sim 0.25$\,dex lower compared to those using Schreiber et al., whereas at $\rm z>1.5$ the two SFR are similar (Table \ref{table_delta_sfr}). We also compare with the MS definition of \cite{Whitaker2014} (their equation 3). There is a small systematic offset at $\rm z>1$. At this redshift interval, SFR measurements of X-CIGALE are lower by $\sim 0.15$\,dex compared to SFR values using the expression of Whitaker et al. (right panel of Fig. \ref{fig_cigale_vs_schreiber}). At $\rm z<1$, the two SFR estimations are similar (distribution peaks at zero). We conclude, that the small offset we observe between X-CIGALE SFR measurements and those using the Schreiber et al. formula, is not a systematic mis-calculation of the SED fitting algorithm but is most likely related to the lack of a rigorous definition of the MS due to the uncertainties and different analysis in the estimations of galaxy properties among the various studies. 

%For example, using equation (3) and the values presented in Table 3 (for $\alpha _{\rm high}$) from \cite{Whitaker2014},

In the next Section, we study the SFR$_{norm}-$L$_X$ relation, when the same methodology (i.e., SED fitting and using the same parametric grid) is applied to measure the SFR and M$_*$ of both X-ray and non X-ray systems and using our own definition of the star forming MS. We also examine how this relation changes, when the systematics and selection effects, presented in this Section, are not taken into account.

%completeness limit that have been excluded from the measurements presented in the bottom panel of the Figure. Furthermore, there are small but systematic offsets of the SFR calculations using X-CIGALE compared to the SFR measurements using the \cite{Schreiber2015} analytic formula, as shown in the left panel of Fig. \ref{fig_cigale_vs_schreiber} and in Table \ref{table_delta_sfr}. 

%At $\rm z<1.5$ SFR values from X-CIGALE are $\sim 0.2$5\,dex lower compared to those using Schreiber et al., whereas at $\rm z>1.5$ the two SFR are similar. We also use different MS definitions from the literature to calculate the SFR of star-forming galaxies and compare it with X-CIGALE measurements. For example, using equation (3) and the values presented in Table 3 (for $\alpha _{\rm high}$) from \cite{Whitaker2014}, there is a small systematic offset at $\rm z>1$. At this redshift interval, SFR measurements of X-CIGALE are lower by $\sim 0.15$\,dex compared to SFR values using the expression of Whitaker et al. (right panel of Fig. \ref{fig_cigale_vs_schreiber}). At $\rm z<1$, the two SFR estimations are similar (distribution peaks at zero). We conclude, that the small offset we observe between X-CIGALE SFR measurements and those using the Schreiber et al. formula, is not a systematic mis-calculation of the SED fitting algorithm but is most likely related to the lack of a rigorous definition of the MS due to the uncertainties and different analysis in the estimations of galaxy properties among the various studies. 

\begin{table}
\caption{Mean ($\mu$) and standard deviation ($\sigma$) of the difference in the calculations of SFR ($\rm \Delta\rm log\,SFR$), using X-CIGALE and the expressions derived in \cite{Schreiber2015} and \cite{Whitaker2014}, for MS star forming galaxies (see also Fig. \ref{fig_cigale_vs_schreiber}).}
\centering
\setlength{\tabcolsep}{1.5mm}
\begin{tabular}{ccc}
\multicolumn{3}{c}{$\rm \Delta\rm log\,SFR$} \\
\hline
redshift range & Schreiber et al. & Whitaker et al. \\
 & $\mu, \sigma$ & $\mu, \sigma$ \\
 \hline
$\rm 0.5<z<1.0$ & -0.24, 0.15 & 0.03, 0.29 \\
$\rm 1.0<z<1.5$ & -0.25, 0.27 & -0.17, 0.28 \\
$\rm 1.5<z<2.0$ & -0.07, 0.32 & -0.15, 0.33 \\
\label{table_delta_sfr}
\end{tabular}
\end{table}

\begin{figure}
\centering
\begin{subfigure}{.5\textwidth}
  \centering
  \includegraphics[width=1.\linewidth, height=7cm]{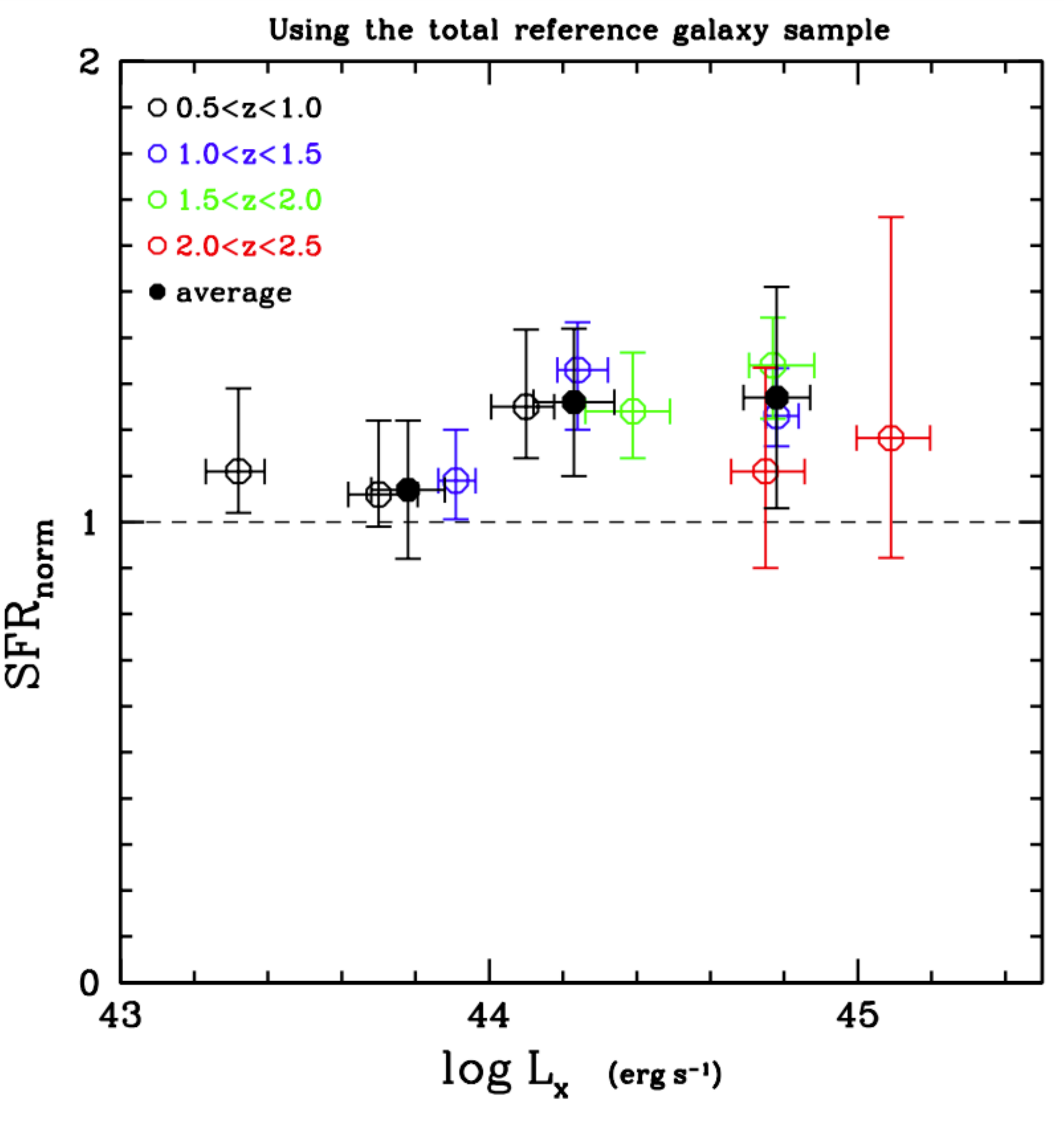}
%  \caption{}
  \label{}
\end{subfigure}
\begin{subfigure}{.5\textwidth}
  \centering
  \includegraphics[width=1.\linewidth, height=7cm]{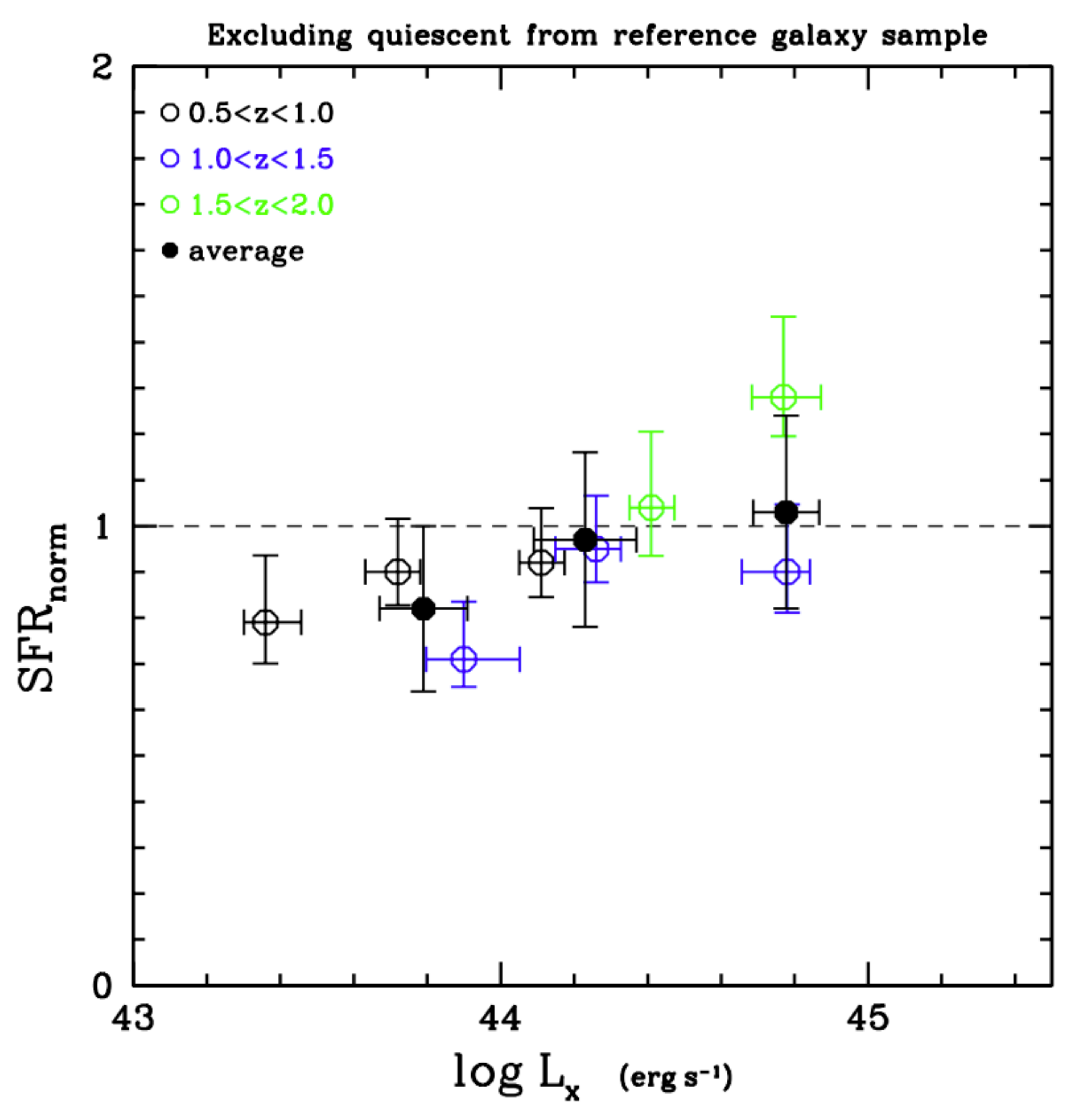}
%  \caption{}
  \label{}
\end{subfigure}
\begin{subfigure}{.5\textwidth}
  \centering
  \includegraphics[width=1.\linewidth, height=7cm]{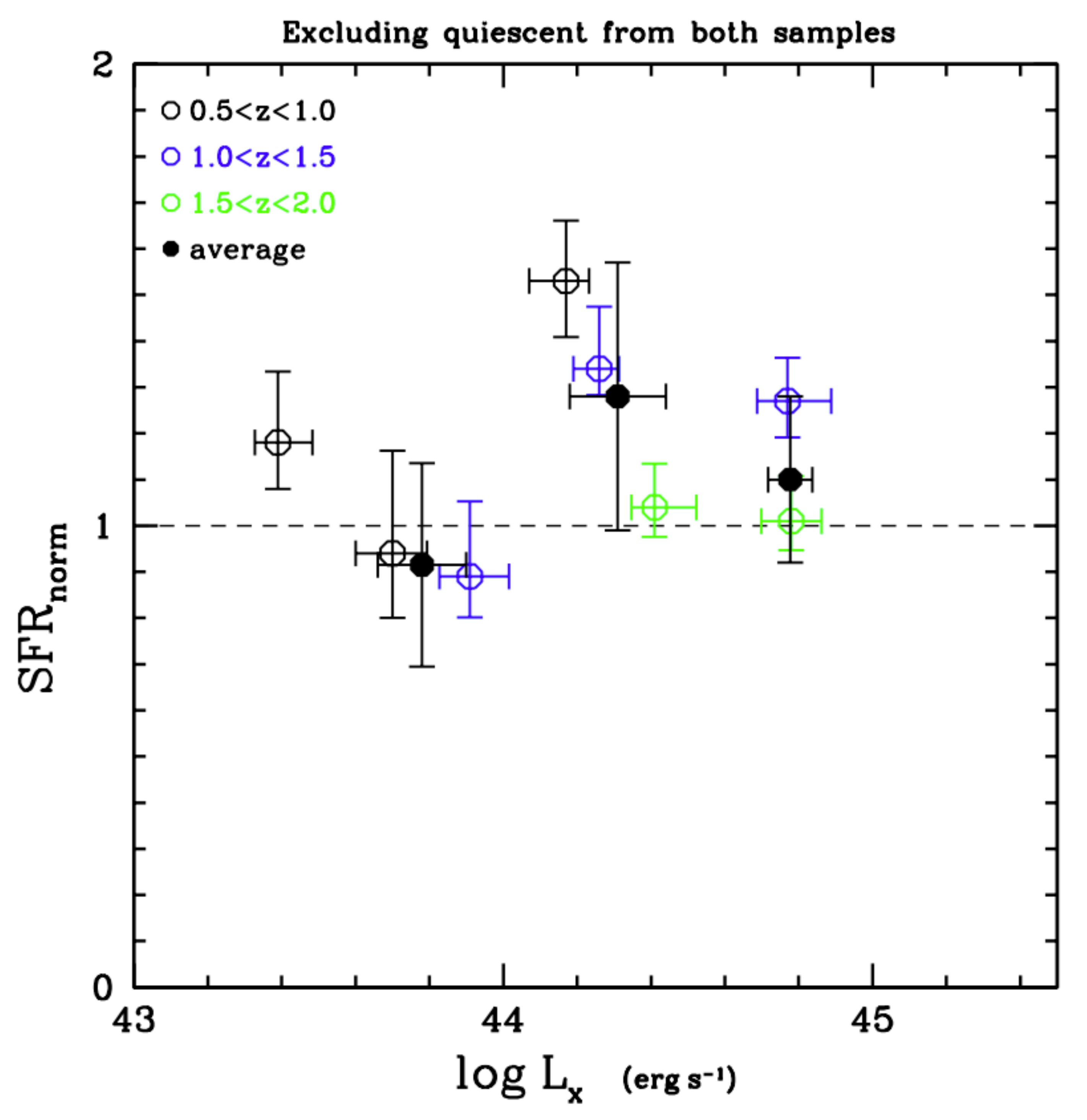}
%  \caption{}
  \label{}
\end{subfigure}
\caption{SFR$_{norm}$ vs. X-ray luminosity. SFR$_{norm}$ and L$_X$ are the median values of our binned measurements, in bins of L$_X$, with 0.5\,dex width. Errors are calculated using bootstrap resampling, by performing 100 resamplings with replacement, at each bin. Results are colour coded based on the redshift interval. Top panel shows the SFR$_{norm}$, estimated using our reference galaxy catalogue with SFR measurements from X-CIGALE. Middle panel presents the results, when we exclude quiescent galaxies from the reference catalogue. Bottom panel shows the measurements when quiescent systems are excluded from both the X-ray and the reference galaxy catalogues (see text for more details).}
\label{fig_sfrnorm_lx}
\end{figure}

\begin{figure}
\centering
  \includegraphics[width=1.\linewidth, height=8.cm]{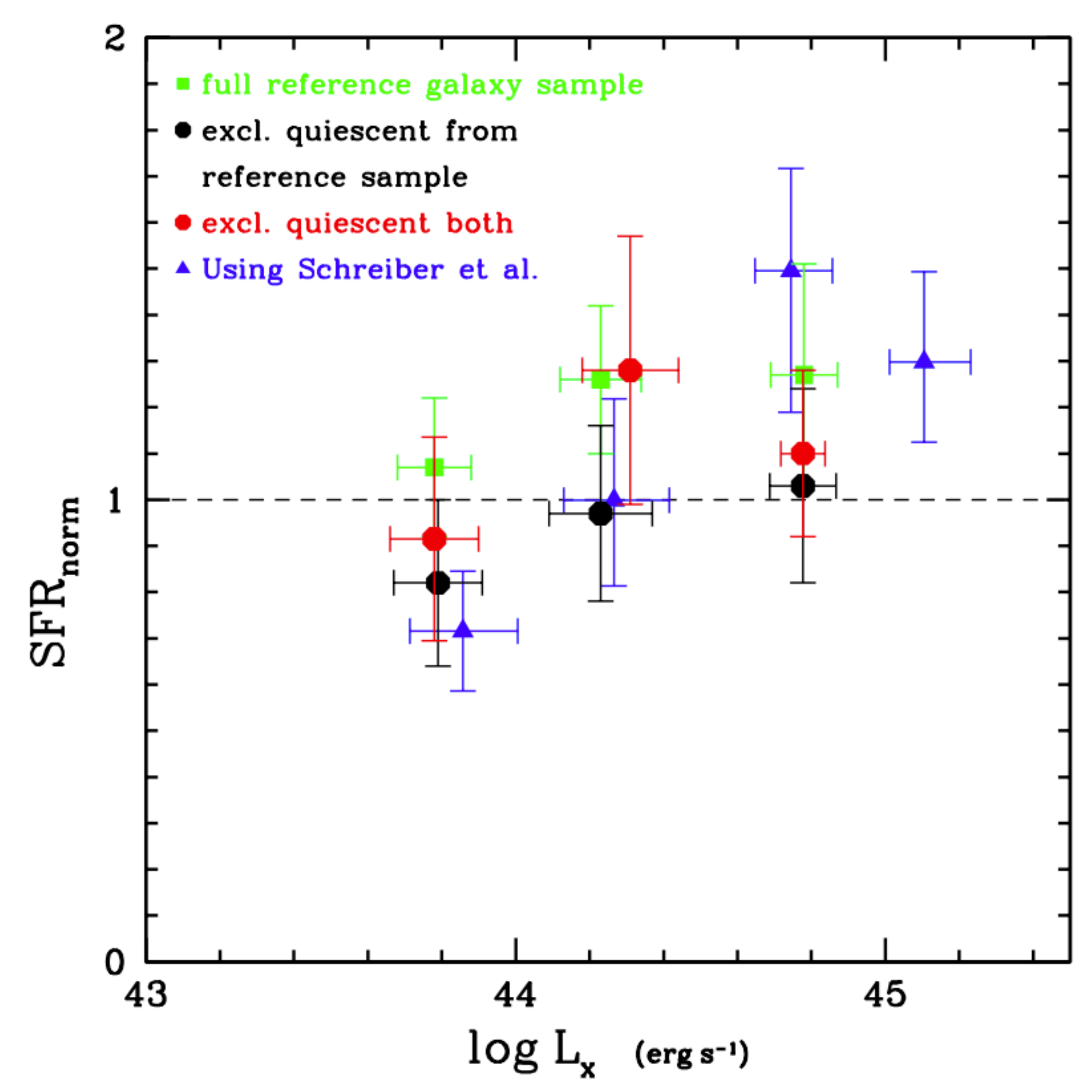}
  \caption{SFR$_{norm}$ vs. X-ray luminosity. SFR$_{norm}$ and L$_X$ are the mean values of the measurements presented in Figures \ref{fig_sfrnorm_lx} and \ref{fig_sfrnorm_lx_schreiber}, grouped in L$_X$ bins of 0.5\,dex, at all redshfits combined, and weighted based on the number of sources in each bin shown in Fig. \ref{fig_sfrnorm_lx}. Errors represent the standard deviation of SFR$_{norm}$ and L$_X$ in each bin.}
  \label{fig_ave_sfrnorm_lx}
\end{figure}

\section{Results I: AGN power and SFR}
\label{sec_lx_sfr}

%However, in our analysis we have used a more refined redshift binning, with narrower redshift intervals. We also suggest that the apparent redshift evolution found in Masoura et al. may hinder dependence of SFR$_{norm}$ on another galaxy property (i.e. stellar mass, see Sections \ref{sec_effect_mstar} and \ref{sec_conclusions}}

\subsection{SFR of X-ray AGN relative to our non X-ray galaxy sample} 
\label{sfrnorm_using_control_sample}

\subsubsection{SFR$_{norm}$ measurements using quiescent and star forming galaxies}

%As noted earlier, most previous works that studied the SFR of X-ray AGN relative to the SFR of MS galaxies, used expressions from the literature to calculate the SFR of MS sources. However, this method hints at a number of systematics and uncertainties. For example, X-ray and non X-ray galaxy samples are defined differently among studies and different methods and/or (SED fitting) algorithms are applied to estimate galaxy properties. The exact definition of MS may also differ among studies.

We use the 18,248 galaxies from our reference galaxy catalogue (see Section \ref{sec_sample_selection}), to estimate SFR$_{norm}$ values for the 1020 X-ray sources that are above the mass completeness limit (Table \ref{table_data}). The SFR of each X-ray AGN is divided by the SFR of galaxies from the reference catalogue that are within $\pm 0.1$\,dex in M$_*$ and $\pm 0.075\times (1+\rm z)$ in redshift. The latter flexible criterion, allows us to increase the number of sources from the reference sample used at high redshifts, thus making our SFR$_{norm}$ calculations more robust. As mentioned, each source is weighted based on the uncertainty on the SFR and M$_*$ parameters (see Section \ref{sec_reliability}). Then, the median values of these ratios is used as the SFR$_{norm}$ of the X-ray source. Only X-ray sources for which their SFR$_{norm}$ has been estimated using at least 30 galaxies from the reference catalogue are used in our measurements. This limit lowers to at least five galaxies for the highest redshift bin, due to the smaller number of available sources. Our measurements are not sensitive to the choice of the box size around the AGN. Changing the above boundaries to $0.05-0.2$ does not change the observed trends, but affects the errors of the calculations. 

The top panel of Fig. \ref{fig_sfrnorm_lx} presents the results of our calculations. Measurements are the median values of SFR$_{norm}$, grouped in L$_X$ bins of size 0.5\,dex. Errors are calculated using bootstrap resampling \citep[e.g.][]{Loh2008}, by performing 100 resamplings with replacement, at each bin. As shown in Table \ref{table_data}, the highest redshift bin includes a small number of sources and thus these measurements should be taken with caution. %SFR$_{norm}$ values at $\rm 1.5<z<2.0$ (green circles) appear lower compared to those at lower redshifts. The same trend can also be seen in the top panel. This redshift bin is dominated by massive galaxies. We discuss further the effect of stellar mass on our measurements in Sections  \ref{sec_effect_mstar} and \ref{sec_conclusions}. Furthermore, 
The SFR$_{norm}$ values using the reference galaxy catalogue present only a mild increase, if any, with X-ray luminosity. The transition point is at $\rm L_{X,2-10keV}\sim10^{44}\,ergs^{-1}$. SFR$_{norm}$-L$_X$ appears flat at higher L$_X$ and redshifts. \cite{Rosario2012}, used X-ray selected AGN from the GOODS-South, GOODS-North and COSMOS fields and found similar results (see their Figure 4). Specifically, their analysis showed that, when mean values are used, SFR (L$_{60}$) of AGN increases with the AGN luminosity at low redshifts with a turning point at $\rm L_{X,2-10keV}\sim10^{44}\,ergs^{-1}$, while at $\rm z>1.5$ they observed a flattening of the relationship of the two parameters for luminous sources.

Filled circles present the weighted average SFR$_{norm}$, in bins of L$_X$, over the total redshift range (also shown with green points in Fig. \ref{fig_ave_sfrnorm_lx}). Mean values are weighted based on the number of sources included in each L$\rm _X$ bin, shown by the coloured circles. This allows us to downweigh measurements from bins with small number of sources (e.g. at $\rm 2.0<z<2.5$). Errors present the standard deviation of the measurements. The SFR of X-ray AGN appears enhanced compared to that of the galaxy sample, by $\sim 20\%$. Although this difference is not constant, but appears lower at $\rm L_{X}<10^{44}\,ergs^{-1}$ ($\sim 10\%$) and higher at $\rm L_{X}>10^{44}\,ergs^{-1}$ ($\sim 30\%$), is systematic across all L$_X$ spanned by our dataset. This can also be seen by the individual points presented in the top panel of Fig. \ref{fig_sfrnorm_lx}. \cite{Florez2020} used 898 X-ray AGN with $\rm L_{X}>10^{44}\,ergs^{-1}$ selected in the Stripe 82 field and compared their SFR with a sample of $\sim 320,000$ non X-ray galaxies. Both of their samples are selected similarly and their galaxy properties have been measured consistently using the same code (CIGALE). They found that X-ray AGN have on average 3-10$\times$ higher SFR than their control galaxy sample, at fixed stellar mass and redshift. However, their calculations are based on mean SFR values. If we consider mean instead of median SFR$_{norm}$ values for the coloured bins and then estimate their weighted average, the SFR of X-ray AGN host galaxies is enhanced by $\sim 40\%$. This is still lower than what Florez et al. claim.

%In Fig. \ref{fig_ave_sfrnorm_lx}, the SFR$_{norm}$ values averaged across all redshift ranges are shown (green squares) and compared to those using the Schreiber et al. formula (blue triangles). SFR$_{norm}$ increases by $\sim 2\times$ from the lowest L$\rm _X$ bin {\bf{up to  $\rm L_{X,2-10keV}\sim 10^{45}\,erg^{-1}$,}} when using Schreiber et al. and by only $\sim 20\%$ using our SFR measurements for the reference galaxy catalogue. {\bf{In the latter case, this mild dependence occurs at $\rm L_{X,2-10keV}\sim10^{44}\,ergs^{-1}$. At  higher X-ray luminosities SFR$_{norm}$ does not seem to correlate with L$_X$.}}

\subsubsection{SFR$_{norm}$ measurements excluding quiescent systems}

%Our analysis in the previous Section showed that, when the Schreiber et al. formula is used to estimate SFR of MS galaxies and then calculate SFR$_{norm}$, we find increase of SFR$_{norm}$ with X-ray luminosity. However, when SFR$_{norm}$ is estimated using our reference control sample, we find a mild(er) evolution with L$_X$ and only for sources at $\rm z<1.5$.

In this Section, we re-calculate the SFR$_{norm}$ of each X-ray AGN, using our own star forming galaxy sample. For that purpose, we use the analysis described in Section \ref{sec_systematics} (see also Fig. \ref{fig_ssfr}) to identify and exclude quiescent galaxies. The results are presented in the middle panel of Fig. \ref{fig_sfrnorm_lx}. SFR$_{norm}$ values appear lower compared to those presented in the previous Section (top panel of Fig. \ref{fig_sfrnorm_lx}). This is expected since we have now excluded quiescent systems from the reference sample. In agreement with our previous calculations, the results show no evolution of SFR$_{norm}$ with redshift. Regarding the dependence of the SFR$_{norm}$ on L$_X$, we confirm our previous findings for a mild dependence. This is also illustrated by the average SFR$_{norm}$ values across all redshifts spanned by our dataset (black circles in Figures \ref{fig_sfrnorm_lx} and \ref{fig_ave_sfrnorm_lx}).

Next, we exclude quiescent systems from both the X-ray and the reference galaxy catalogues. Based on our definition of quiescent systems, $\sim 25\%$ of X-ray AGN live in quiescent galaxies. The percentage is higher at $\rm z<1$ ($\sim 35\%$) and drops to $\sim 7\%$ at $\rm z>1.5$.  These percentages are similar to those found for the galaxies in the reference sample in Section \ref{sec_systematics}. In the bottom panel of Fig. \ref{fig_sfrnorm_lx}, we present the SFR$_{norm}$ values as a function of X-ray luminosity. Filled circles show the weighted average SFR$_{norm}$ values estimated in L$_X$ bins of width 0.5\,dex and weighted based on the number of sources in the bins, shown by the open circles. As expected, SFR$_{norm}$ values are higher compared to those when we exclude quiescent systems only from the reference sample, however they follow the same trends with our previous measurements. Specifically, we do not find dependence of SFR$_{norm}$ with redshift while there is a mild increase ($\sim 30\%$) of SFR$_{norm}$ with L$_X$, only at $\rm L_{X,2-10keV}<10^{44.5}\,ergs^{-1}$.

%\begin{figure}
%\centering
%  \includegraphics[width=1.\linewidth, height=7.cm]{sfrnorm_average_Lx_excl_quiesc_both_frac_v2_ref.pdf}
%  \caption{SFR$_{norm}$ vs. X-ray luminosity. Quiescent systems have been excluded from both the X-ray and the reference galaxy catalogues (see text for more details). The results are colour coded based on the redshift interval. Median values are presented and the errors are calculated using bootstrap resampling. Black squares present the mean SFR$_{norm}$ values grouped in L$_X$ bins of 0.5\,dex and weighted based on the number of sources included in each L$_X$ bin (i.e. coloured circles), accross all redshifts. The errors on the average values correspond to the standard deviation.}
%  \label{fig_ave_sfrnorm_lx_excl_quiesc_both}
%\end{figure}

\subsection{SFR of X-ray AGN relative to main sequence galaxies, defined  in previous studies} 
\label{sec_using_schreiber}

%In this Section, we examine whether there is a connection between the AGN power and the SFR of the host galaxy. The SFR of X-ray AGN evolves with stellar mass and redshift in a way that is qualitatively similar to the SFR evolution of star-forming galaxies \citep[Fig. 7 in][]{Masoura2018}. To account for this evolution, previous studies estimate the SFR$_{norm}$ parameter, that is defined as the ratio of the SFR of X-ray AGN to the SFR of star-forming MS galaxies, $\rm SFR_{{\it{norm}}}=\frac{SFR_{AGN}}{SFR_{MS}}$, \citep{Mullaney2015, Masoura2018, Bernhard2019, Aird2019, Grimmett2020, Masoura2021}. For that purpose, nearly all of them, use the analytical expression (9) of \cite{Schreiber2015} to parametrize the SFR of main-sequence galaxies. 

Recent studies that examined the correlation of SFR$_{norm}$ with L$_X$, used equation 9 of Schreiber et al. to calculate SFR$_{norm}$. In this Section, we follow their approach. Our goal is to compare our findings following this methodology with their results and most importantly with our measurements presented in the previous Section.

\begin{figure}
\centering
  \centering
  \includegraphics[width=1.\linewidth, height=7cm]{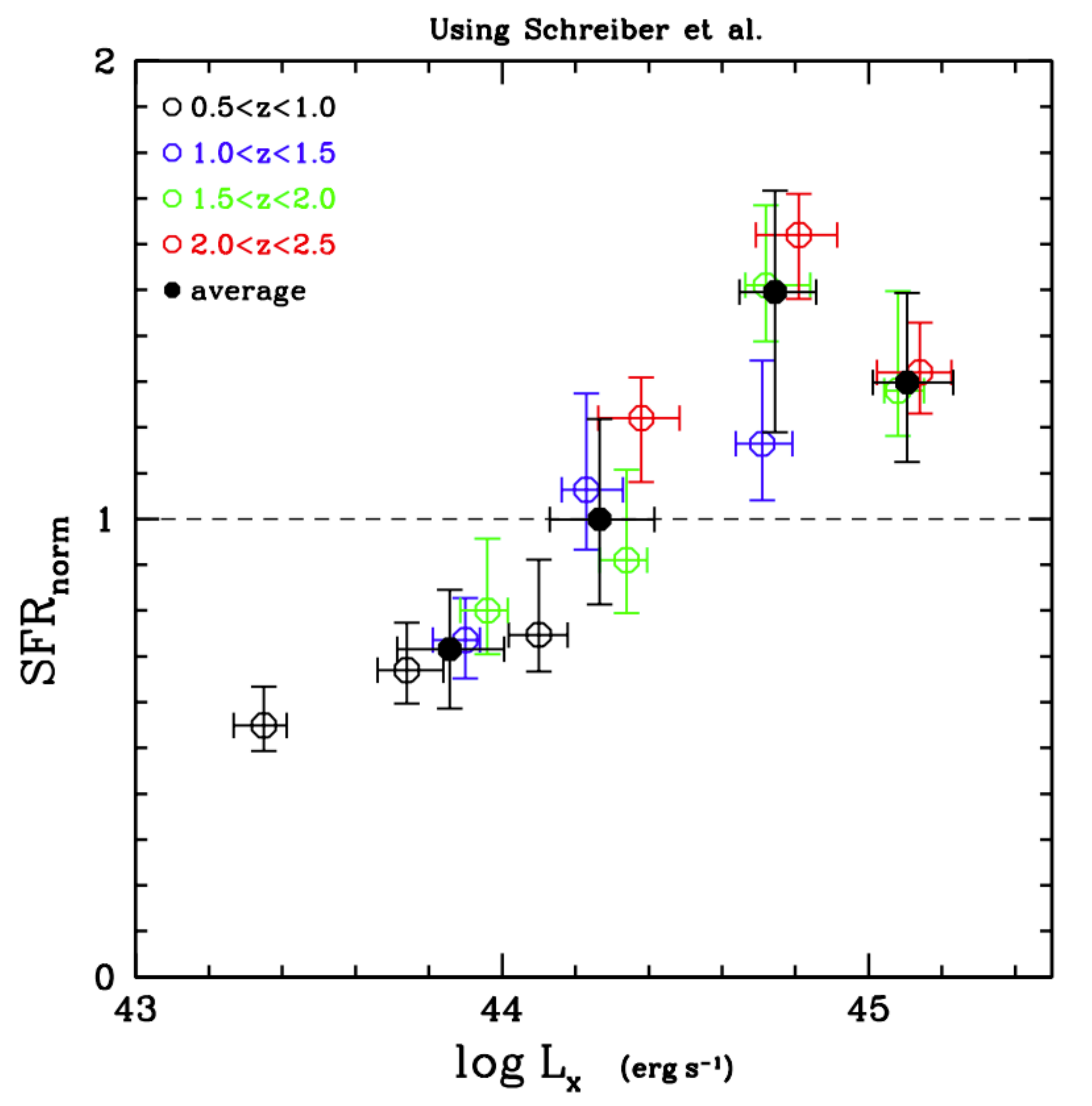}
%  \caption{}
  \label{}
\caption{SFR$_{norm}$ vs. X-ray luminosity. SFR$_{norm}$ values are estimated using the expression of \cite{Schreiber2015} to measure the SFR of star forming MS galaxies. SFR$_{norm}$ and L$_X$ are the median values of our binned measurements, in bins of L$_X$, with 0.5\,dex width. Errors are calculated using bootstrap resampling, by performing 100 resamplings with replacement, at each bin. Results are colour coded based on the redshift interval.}
\label{fig_sfrnorm_lx_schreiber}
\end{figure}

In Fig. \ref{fig_sfrnorm_lx_schreiber}, we plot the SFR$_{norm}$, estimated using the equation 9 of Schreiber et al., as a function of X-ray luminosity, in four redshift bins. For these measurements, we have used the total X-ray sample, i.e. without excluding sources below the mass completeness limit (see Section \ref{sec_mass_completeness}). This is to allow a direct comparison with previous studies that followed the same approach. Measurements are the median values of SFR$_{norm}$, grouped in L$_X$ bins of size 0.5\,dex. Errors are calculated using bootstrap resampling. Each source is weighted based on the uncertainty on the SFR and M$_*$ parameters (see Section \ref{sec_reliability}). SFR$_{norm}$ increases with L$\rm _X$ at all redshift intervals, up to X-ray luminosities $\rm L_{X,2-10keV}<10^{45}\,ergs^{-1}$. At higher luminosities, SFR$_{norm}$ values  appear lower. However, the two bins, at the highest L$_X$ regime spanned by our sample, include a very small number of X-ray sources ($<30$ sources). We do not detect evolution of SFR$_{norn}$ with redshift. Filled circles present the mean SFR$_{norm}$ in bins of L$\rm _X$, over the total redshift range studied in this work ($\rm 0.5<z<2.5$). The width of each bin is 0.5\,dex. Mean values are weighted based on the number of sources included in each L$\rm _X$ bin shown by the coloured circles. Errors present the standard deviation of the measurements. Based on these results, SFR$_{norm}$ presents a strong evolution with L$_X$. Specifically, SFR$_{norm}$ increases by a factor of $\approx 2$ up to $\rm L_{X,2-10keV}<10^{45}\,ergs^{-1}$.

\subsubsection{Comparison with previous studies}

\cite{Mullaney2015}, used 110 AGN at $\rm 0.5<z<1.5$ selected from {{\it{Chandra}} Deep Field North (CDFN) and South (CDFS) and 49 AGN at $\rm 1.5<z<4.0$ from CDFS. They compare the SFR of their AGN sample with the SFR of MS galaxies, using for the latter the equation of Schreiber et al. They found that the SFR distributions of their AGN and MS galaxy samples differ. However, AGN have mean SFR$_{norm}\sim 1$. They attributed the apparent contradiction between the SFR distributions and SFR$_{norm}$ values, to bright outliers that skew their mean SFR$_{norm}$ calculation to higher values. They also found no evolution of the SFR$_{norm}$ between their low and high redshift AGN samples. Using mean values for our calculations, instead of median does not practically affect the L$_X$ values of each bin, but increases the SFR$_{norm}$ values by 0.25-0.50. We also find no evolution of SFR$_{norm}$ with redshift, in agreement with Mullaney et al.

\cite{Bernhard2019}, used 541 AGN in the COSMOS field, within $\rm 0.8<z<1.2$ and found that the SFR$_{norm}$ of higher L$_X$ AGN ($\rm L_{X,2-10keV}>2\times 10^{43}\,ergs^{-1}$) is narrower and closer to that of the MS galaxies than that of lower L$_X$. Although, their sample spans significantly lower X-ray luminosities and lacks high L$_X$ compared to the X-ray sample used in this study (see their Fig.1 and our Fig. \ref{fig_redz_lx}), our results are in broad agreement. As shown in Fig. \ref{fig_sfrnorm_lx_schreiber}, our lowest L$_X$ bin lies below the dashed line, i.e. SFR of low L$_X$ AGN is lower than that of MS galaxies, while at $\rm L_{X,2-10keV} \sim 10^{44}\,ergs^{-1}$, the SFR of AGN is consistent with that of MS galaxies (SFR$_{norm} \sim 1$).

The L$_X$ and redshift distributions of our sample is closer to that used in \cite{Masoura2018} and \cite{Masoura2021}. In the latter work, they used 3213 X-ray AGN in the {\it{XMM-XXL}} and found evolution of the SFR$_{norm}$ parameter with X-ray luminosity. Our results are in agreement with their measurements. However, \cite{Masoura2021} find higher SFR$_{norm}$ values for sources at $\rm z>1.2$ compared to those at $\rm z<1.2$ (see their Fig. 8). As already mentioned, we do not find evolution of SFR$_{norm}$ with redshift. We note, that in Masoura et al. the low redshift bin also includes sources at $\rm z<0.5$. Such sources have been excluded from our analysis (and Mullaney et al.), for the reasons mentioned in Section \ref{sec_data}. Perhaps, the redshift evolution of SFR$_{norm}$ found in Masoura et al. is driven by sources at $\rm z<0.5$ and this could be the reason why we (and Mullaney et al.) do not detect it in our samples.

%As already mentioned, we do not find evolution of SFR$_{norm}$ with redshift. We note, that in Masoura et al. the redshift bins were defined differently and that their low redshift bin also includes sources at $\rm z<0.5$. Such sources have been excluded from our analysis, for the reasons mentioned in Section \ref{sec_data}. If we split our sources into two redshift bins, i.e. $\rm 0.5<z<1.2$ and $\rm z>1.2$, we find that the median value of SFR$_{norm}$ is higher by a factor of $\sim 2$ at the high redshift bin compared to the low redshift bin. This factor is lower compared to that found in Masoura et al. ($\sim 10$), but could perhaps increase if we also included sources at $\rm z<0.5$. A better comparison between the two samples can be perfomed by comparing our results with those presented in the left panel of their Fig. 11. In this case, SFR$_{norm}$, between sources at $\rm 0.5<z<1.2$ and $\rm z>1.2$ (their blue and red symbols) increases by a factor of $\sim 2-3$. We conclude that if SFR$_{norm}$ evolves with redshift, this evolution is mostly driven by sources at $\rm z<0.5$ and this could be the reason why we (and Mullaney et al.) do not detect it in our samples. 

\subsubsection{The effect of systematics and selection effects on the SFR$_{norm}-$L$_X$ relation}

Blue triangles in Fig. \ref{fig_ave_sfrnorm_lx}, present the weighted mean values of SFR$_{norm}$ when we are using the Schreiber et al. equation and are compared with the results following our analysis, as presented in Section \ref{sfrnorm_using_control_sample}. We notice that SFR$_{norm}$ values are higher using our definition of MS galaxies, at low redshift and X-ray luminosities ($\rm z<1$, $\rm L_{X,2-10keV}<10^{44}\,erg^{-1}$) and get lower at higher redshift and luminosities, compared to SFR$_{norm}$ calculations using the Schreiber et al. formula. Part of this difference can be attributed to two main factors. First, the X-ray sample used when SFR$_{norm}$ is estimated with the Schreiber formula, includes sources below the mass completeness limit that have been excluded from the measurements presented in the bottom panel of the Figure. Furthermore, there are small but systematic offsets of the SFR calculations using X-CIGALE compared to the SFR measurements using the \cite{Schreiber2015} analytic formula, as shown in the left panel of Fig. \ref{fig_cigale_vs_schreiber} and in Table \ref{table_delta_sfr}. Moreover, at $\rm z<1.5$, SFR$_{norm}$ values using the reference galaxy catalogue are higher compared to those using Schreiber et al. analytical form (top panel of Fig. \ref{fig_sfrnorm_lx}). This is expected, since our reference galaxy catalogue includes a mix of star forming and quiescent galaxies. Schreiber et al. formula estimates the SFR of galaxies in the locus of the main sequence. 

We conclude, that following the same methodology with recent studies to calculate the SFR$_{norm}$ parameter and study its dependence on L$_X$ and redshift, gives results that are generally in agreement with those reported in previous works. However, the results presented in Fig. \ref{fig_ave_sfrnorm_lx}, highlight the importance of studying the SFR$_{norm}-$L$_X$ relation, in a uniform manner, taking into account the systematics and selection effects. Inconsistencies among different methodologies may lead to incorrect conclusions regarding the SFR$_{norm}-$L$_X$ relation.

\begin{figure}
\centering
  \includegraphics[width=1.\linewidth, height=7cm]{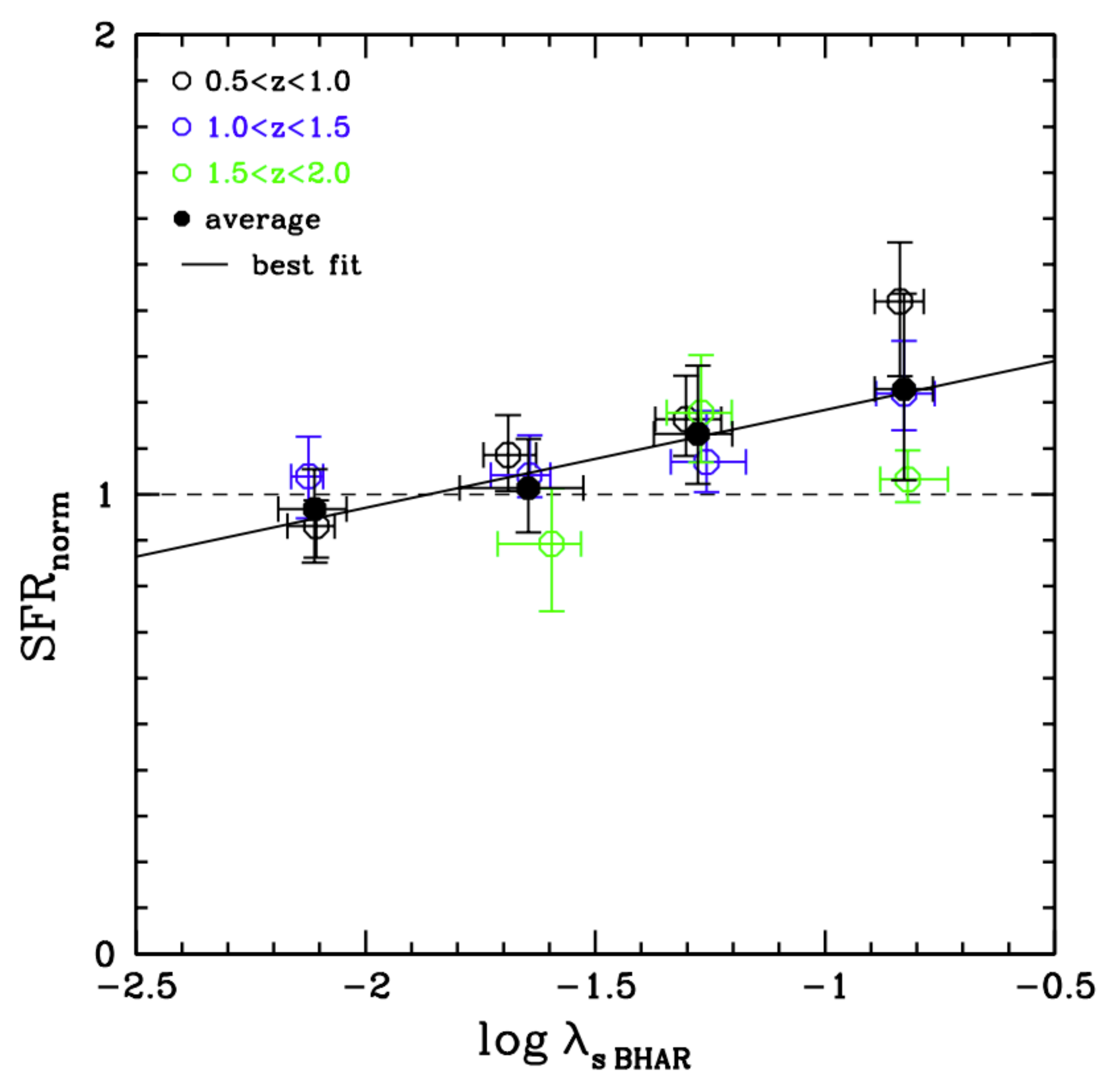}
  \caption{SFR$_{norm}$ vs. specific black hole accretion rate ($\lambda _{s, BHAR}$). SFR$_{norm}$ has been estimated, after excluding quiescent systems from both the X-ray and the reference galaxy catalogues (see text for more details). Results are colour-coded based on the redshift interval. Measurements are grouped in $\lambda _{s, BHAR}$ bins, with 0.5 width. Median values are presented and the errors are calculated using bootstrap resampling. The average SFR$_{norm}$ values (black squares) are weighted based on the number of sources in each bin (coloured circles), accross all redshift intervals. Errors on the average values correspond to the standard deviation within each $\lambda _{s, BHAR}$ bin. Solid line present the best fit of the average values, SFR$_{norm}=\rm 0.213^{+0.035}_{-0.031}log\,\lambda _{s, BHAR}+1.397^{+0.175}_{-0.158}$.}
  \label{fig_sfrnorm_lambda}
\end{figure}

\begin{table*}
\caption{Weighted median values of M$_*$, L$_X$ and redshift in each of the L$_X$ and $\lambda _{s\,BHAR}$ bins, shown in the bottom panel of Figure \ref{fig_sfrnorm_lx}  and in Fig. \ref{fig_sfrnorm_lambda}. The weight is based on the accuracy of the SFR and M$_*$ calculations of X-CIGALE, for each source (see text for more details). }
\centering
\setlength{\tabcolsep}{2mm}
\begin{tabular}{cccccccccccc}
 \hline
& \multicolumn{4}{c}{$\rm log\,\lambda _{s\,BHAR}$} & &&&&\multicolumn{3}{c}{$\rm log\,(L_{X,2-10keV}/(ergs^{-1}))$}  \\
  \hline
& -2.11 & -1.65 & -1.28 & -0.83  & &&&&43.78 & 44.31 & 44.78 \\
  \hline
$\rm log\,[M_*(M_\odot)]$ & 11.45 & 11.36& 11.33 & 11.27  & &&&&11.22 & 11.47 & 11.62 \\
$\rm log\,(L_{X,2-10keV}/(ergs^{-1}))$ &43.27  & 43.65 & 44.07 & 44.64  &&&&& 43.78 & 44.31 &44.78   \\
redshift & 0.79 & 0.97& 1.17 & 1.33  & &&&&0.94 & 1.24 & 1.58 \\
  \hline

\label{table_lx_mstar_redz}
\end{tabular}
\end{table*}

\subsection{Discussion}

{The analysis using our definition of MS galaxies showed only a mild increase of SFR$_{norm}$ with L$_X$,  with a transition luminosity at $\rm L_{X,2-10keV}\sim 10^{44}\,ergs^{-1}$ (Section \ref{sfrnorm_using_control_sample}).  Now, we check whether this result holds when we account for the stellar mass of the examined systems. 

\subsubsection{The effect of stellar mass}
\label{sec_effect_mstar}

To examine if the SFR$_{norm}$-L$_X$ relation differs with stellar mass, we use the specific black hole accretion rate parameter \citep[$\lambda_{s,BHAR}$;][]{Aird2012, Aird2018}, i.e., the rate of accretion onto the supermassive black hole relative to the stellar mass of the host galaxy. If, instead, we use the specific X-ray luminosity \citep{Yang2018} does not change the observed trends and our conclusions. For the estimation of $\lambda_{s,BHAR}$ we use the mathematical expression (2) of \cite{Aird2018}: 

\begin{equation}
\lambda_{s,BHAR}=\frac{k_{bol} \times L_X}{1.3\times 10^{38}\rm {erg\,s^{-1}}\times 0.002\frac{M_*}{M\odot}},
\end{equation}
where $k_{bol}$ is a bolometric correction factor. We adopt the same value as in \cite{Aird2018}, i.e. $k_{bol}=25$. Each source is weighted based on the uncertainty of M$_*$  (see Section \ref{sec_reliability}). The results are shown in Fig. \ref{fig_sfrnorm_lambda}. For these measurements, we have used SFR$_{norm}$ values calculated after excluding quiescent systems from the X-ray and reference galaxy catalogues. Black circles present the average measurements over the total redshift range, in bins of $\lambda_{s,BHAR}$, weighted by the number of sources in each bin, shown by the coloured, open circles. Based on our results, SFR$_{norm}$ increases with the specific accretion rate of the supermassive black hole, by a factor of $\sim 1.3$.

\subsubsection{Are AGN accretion and star formation linked?}
\label{sec_conclusions}

Our analysis shows enhancement of SFR$_{norm}$ with $\lambda_{s,BHAR}$, at all redshifts and specific black hole accretion rates, spanned by our sample. However, the SFR$_{norm}$-L$_X$  relation does not present a consistent trend. In Table \ref{table_lx_mstar_redz}, we present the weighted median values of L$_X$, M$_*$ and redshift for each L$_X$ and $\lambda_{s,BHAR}$ bins, presented in the bottom panel of Fig. \ref{fig_sfrnorm_lx} and in Fig. \ref{fig_sfrnorm_lambda}. The weight is based on the accuracy of the SFR and M$_*$ calculation of X-CIGALE for each source. L$_X$ and $\lambda_{s,BHAR}$ are binned the same across the redshift range spanned by our dataset. Additionally, our previous results showed no evolution of SFR$_{norm}$ with redshift. Therefore, redshift is not a differentiating factor. We notice that $\lambda_{s,BHAR}$ bins of low value, tend to include, on average, the most massive and least luminous systems. On the other hand, when SFR$_{norm}$ is grouped in L$_X$ bins, the most massive systems are included in the highest L$_X$ bins. In Fig \ref{fig_sfrnorm_lx_mstar}, we repeat the measurements presented in the bottom panel of Fig. \ref{fig_sfrnorm_lx}, but at each redshift range, the results are also grouped in stellar mass bins (additionally to L$_X$ bins). The results are colour coded based on the redshift range (black for $\rm 0.5<z<1.0$, blue for $\rm 1.0<z<1.5$ and green for $\rm 1.5<z<2.0$). Different symbols refer to different stellar mass intervals. We notice that for the less massive sources ($\rm log\,[M_*(M_\odot)] \sim 11-11.5$), SFR$_{norm}$ increases with a turning point at $\rm L_{X,2-10keV} \sim 10^{44}\,erg^{-1}$, by a factor of up to $\sim 1.5$. These results, in conjunction with those presented in Fig. \ref{fig_sfrnorm_lambda} and the numbers shown in Table \ref{table_lx_mstar_redz}, may indicate that in less massive systems ($\rm log\,[M_*(M_\odot)] < 11.5$), SFR$_{norm}$ increases with the AGN power, for luminosities up to a few times $\rm 10^{44}\,ergs^{-1}$, while at higher L$_X$ the samples are dominated by massive systems in which the SFR$_{norm}$-L$_X$ relation appears flat. We note, though, that a strong conclusion cannot be drawn, as the trend is based on bins that include a small number of AGN ($\sim 120$). Furthermore, data from deeper surveys that will provide us with AGN at lower luminosities are needed to provide corroborating evidence.

%It would also be interesting to examine whether the SFR$_{norm}$-L$_X$ relation of very massive systems, at lower L$_X$ appears flat.}}  

%This implies that SFR$_{norm}$ increases with L$_X$ regardless of the mass of the host galaxy, i.e. the power (L$_X$) of X-ray AGN enhances the SFR of the host galaxy relative to the SFR of a non-AGN star forming galaxy, regardless of the size (stellar mass) of the galaxy it lives in.

\begin{figure}
\centering
  \includegraphics[width=1.\linewidth, height=7cm]{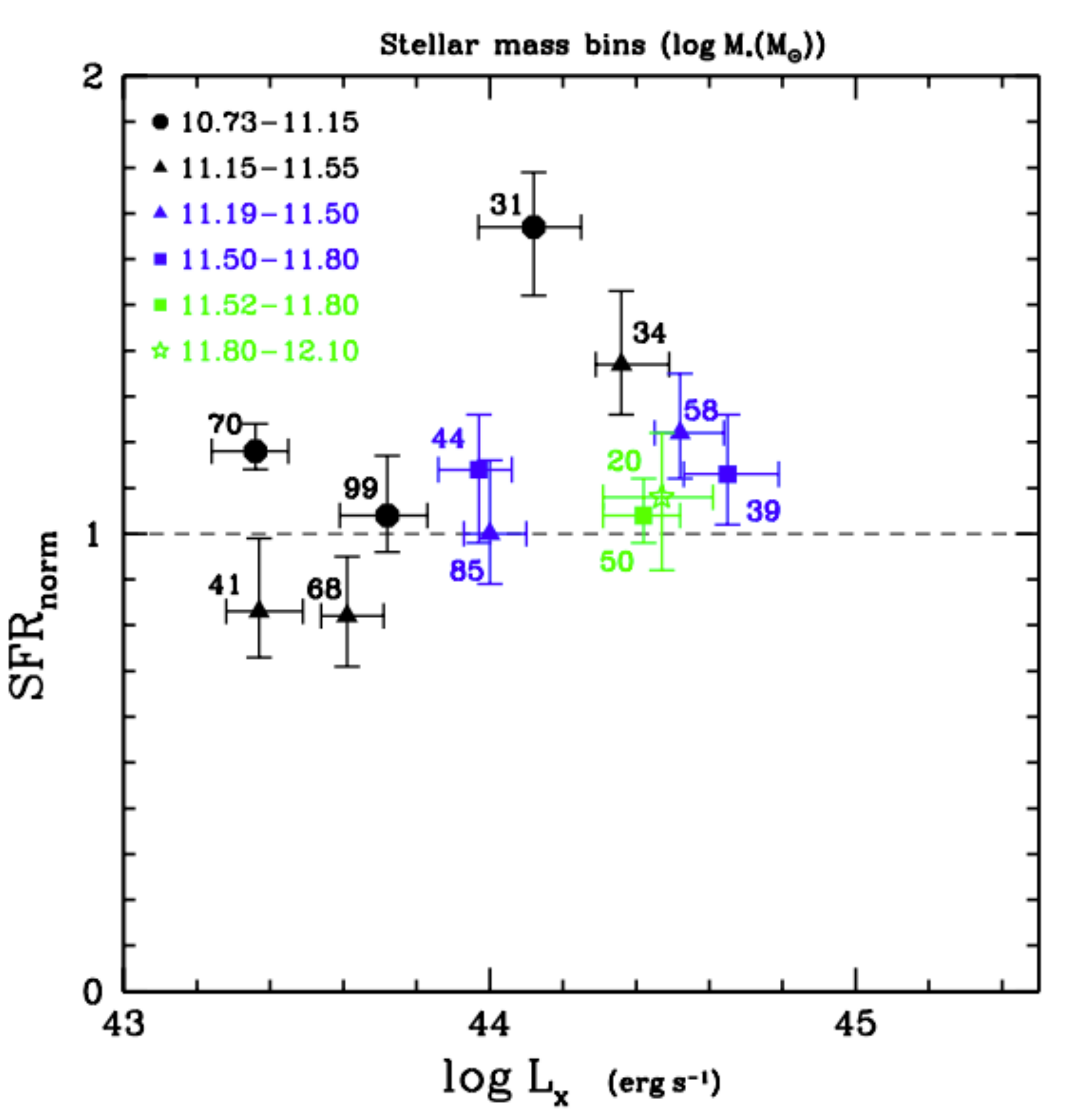}
  \caption{SFR$_{norm}$ vs. X-ray luminosity. SFR$_{norm}$ has been estimated, after excluding quiescent systems from both the X-ray and the reference galaxy catalogues (see text for more details). Results are grouped in L$_X$ bins of 0.5\,dex width and stellar mass bins. The results are colour coded based on the redshift range (black for $\rm 0.5<z<1.0$, blue for $\rm 1.0<z<1.5$ and green for $\rm 1.5<z<2.0$). Different symbols refer to different stellar mass intervals, as indicated in the legend. M$_*$ values are in $\rm log\,[M_*(M_\odot)]$. Median SFR$_{norm}$ values are presented and the errors are calculated using bootstrap resampling. The number of sources included in each bin are shown next to each symbol.}
  \label{fig_sfrnorm_lx_mstar}
\end{figure}

Although the results appear marginally significant, it is interesting to consider their plausible interpretations, if we assume that there is a correlation between the SMBH accretion and galaxy star formation and this correlation exists only in lower mass systems ($\rm log\,[M_*(M_\odot)] < 11.5$). Instantaneous AGN accretion does not appear to track star formation in the most massive galaxies ($\rm log\,[M_*(M_\odot)] > 11.5$) and at high accretion rates ( $\rm L_{X,2-10keV}>10^{44.5}\,erg^{-1}$). A scenario that could explain this hypothesis is that different physical mechanisms fuel AGN at these two mass regimes. Indeed, AGN constitute a diverse population and previous works, both observational \citep[e.g.][]{Allevato2011, Mountrichas2013, Mountrichas2016} and theoretical \citep[e.g.][]{Fanidakis2012, Fanidakis2013} found that different AGN triggering processes may be dominant depending on e.g. redshift, X-ray luminosity and mass of the source. 

%These indicate that AGN feedback, e.g. outflows of ionized gas \citep[e.g.][]{Revalski2021}, may enhance the star formation of the host galaxy in less massive systems ($\rm log\,[M_*(M_\odot)] \sim 11$), but it does not impact the star formation in more massive galaxies. 

\begin{figure*}
\centering
\begin{subfigure}{.505\textwidth}
  \centering
  \includegraphics[width=.92\linewidth, height=6.2cm]{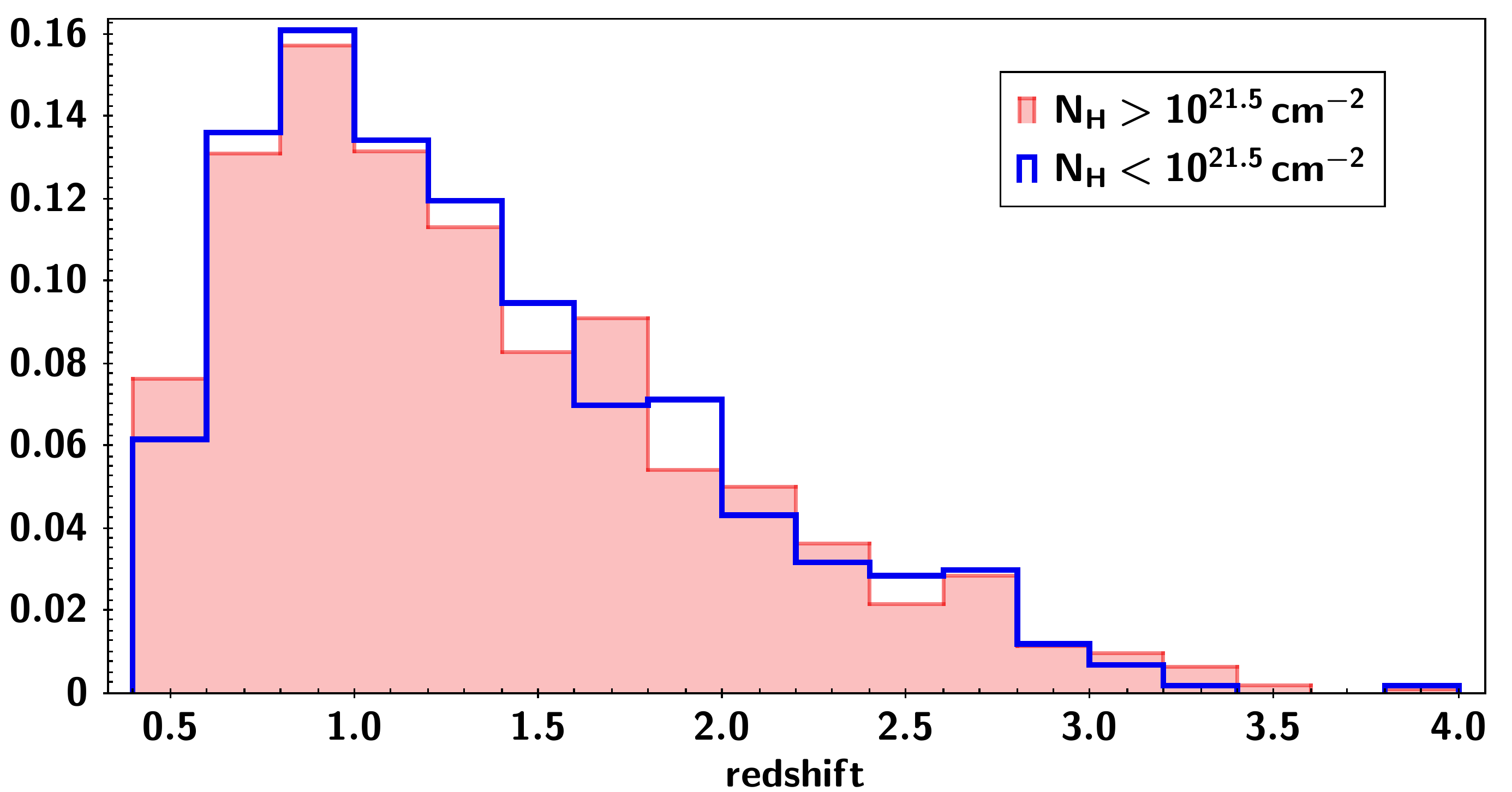}
%  \caption{}
  \label{z}
\end{subfigure}%
\begin{subfigure}{.52\textwidth}
  \centering
  \includegraphics[width=.92\linewidth, height=6.2cm]{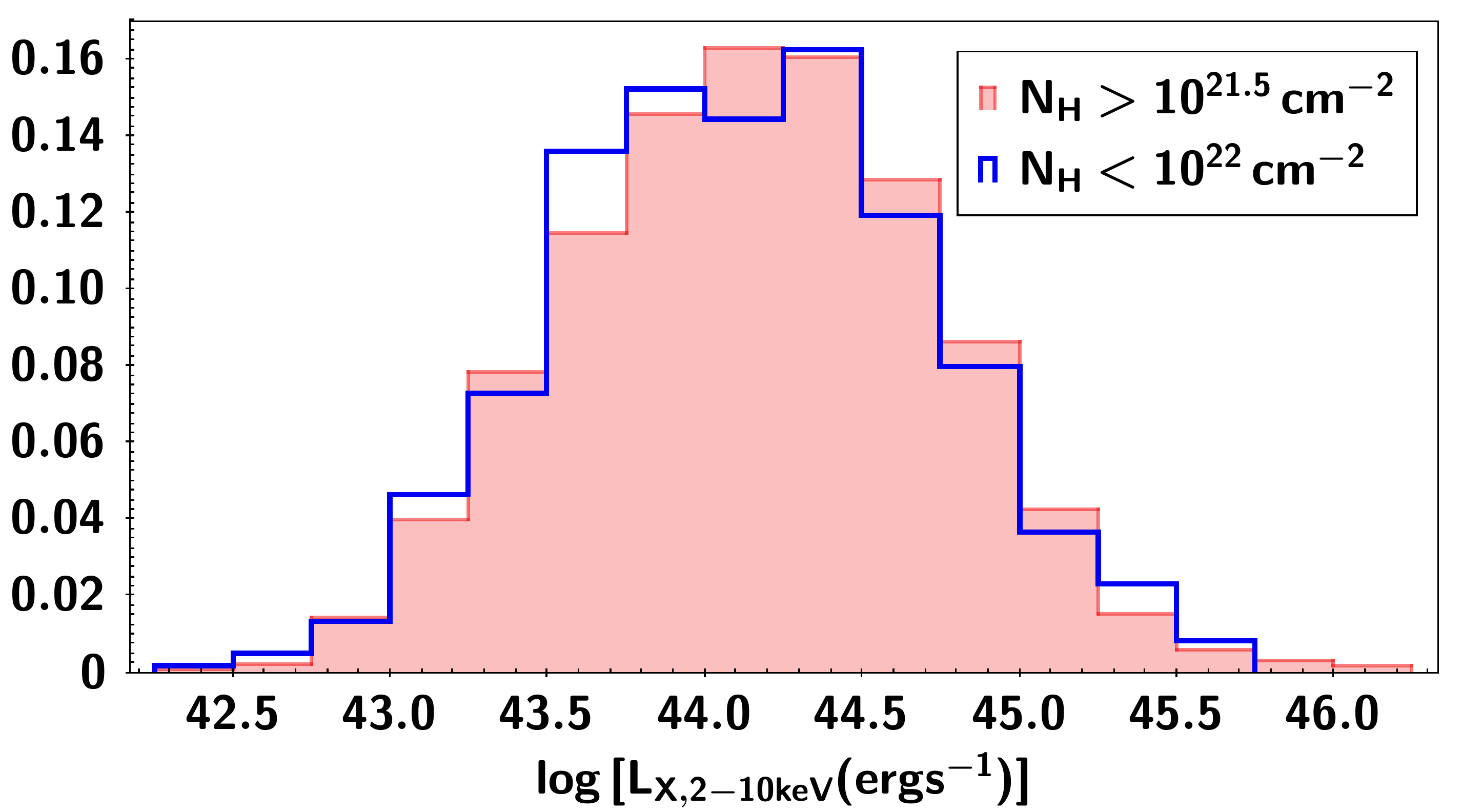}
%  \caption{}
  \label{Lx}
\end{subfigure}
\caption{Left: Redshift distribution. Right: Distribution of the X-ray hard, intrinsic luminosity. Blue and red histograms refer to the X-ray unabsorbed and absorbed sources, respectively. Both histograms have been normalised to the total number of sources.}
\label{fig_z_Lx_xray}
\end{figure*}

\begin{figure}
\centering
\begin{subfigure}{.5\textwidth}
  \centering
  \includegraphics[width=1.\linewidth, height=5.2cm]{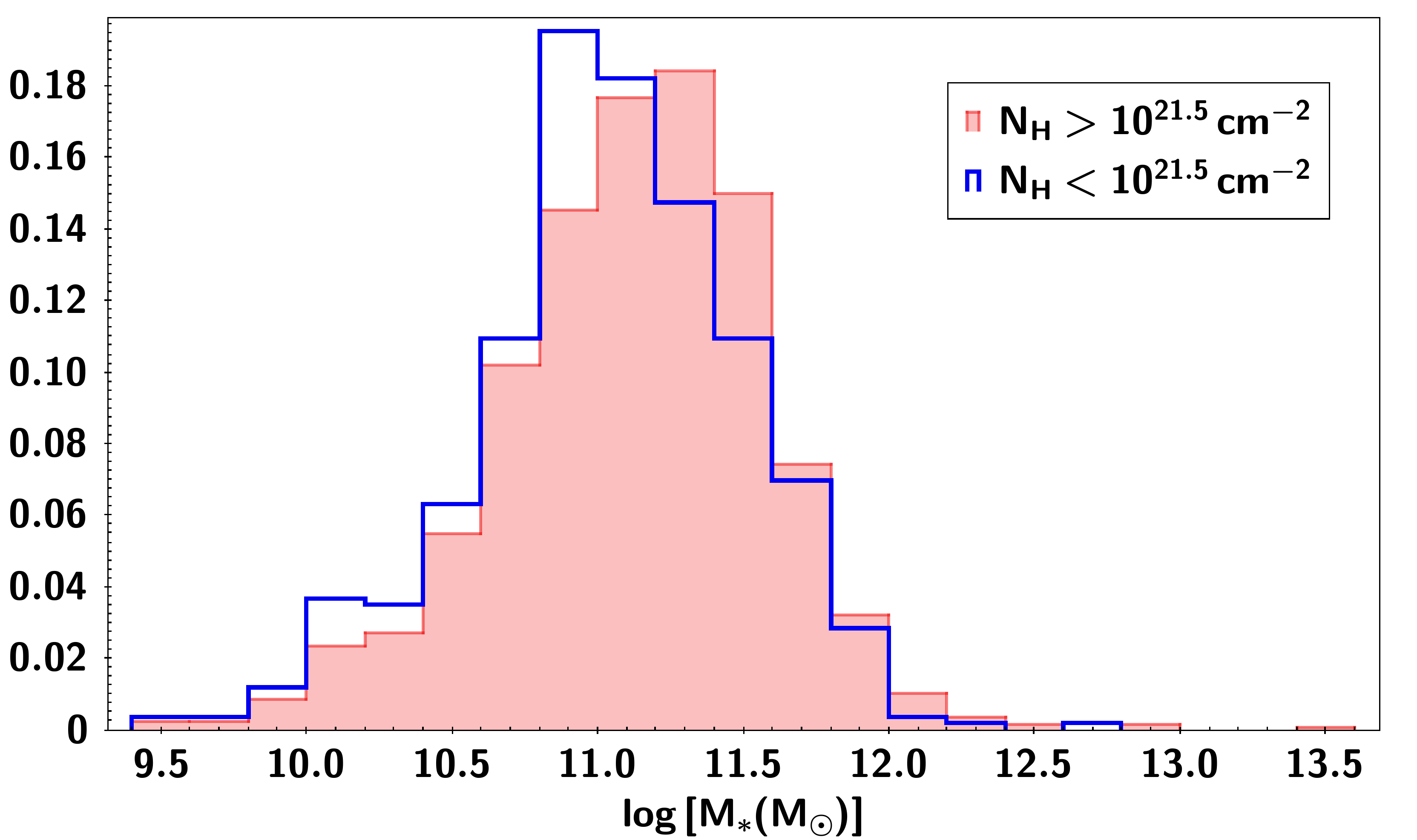}
%  \caption{}
  \label{fig_mstellar_xray}
\end{subfigure}
\begin{subfigure}{.5\textwidth}
  \centering
  \includegraphics[width=1.\linewidth, height=5.2cm]{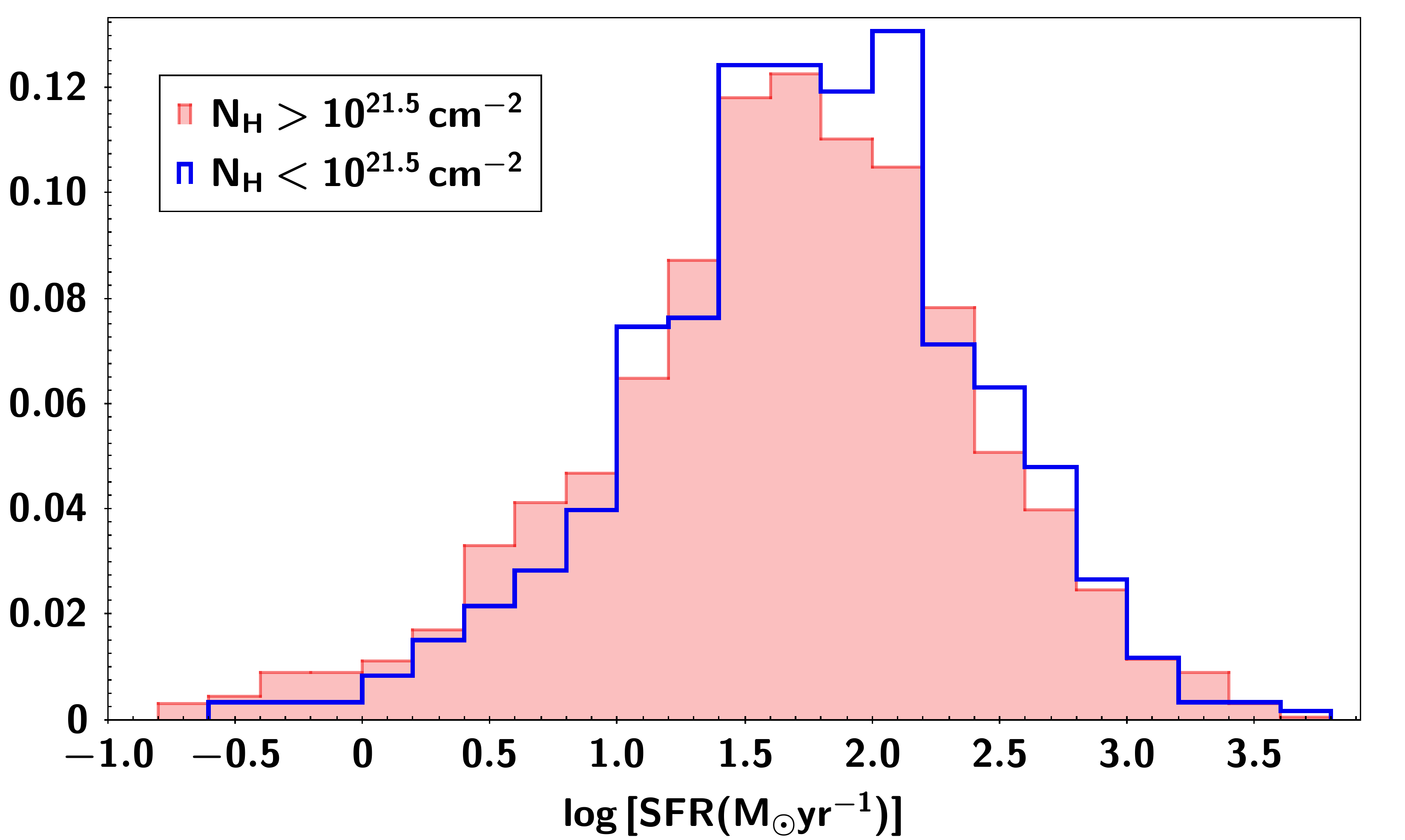}
%  \caption{}
  \label{fig_sfr_xray}
\end{subfigure}
\begin{subfigure}{.5\textwidth}
  \centering
  \includegraphics[width=1.\linewidth, height=5.2cm]{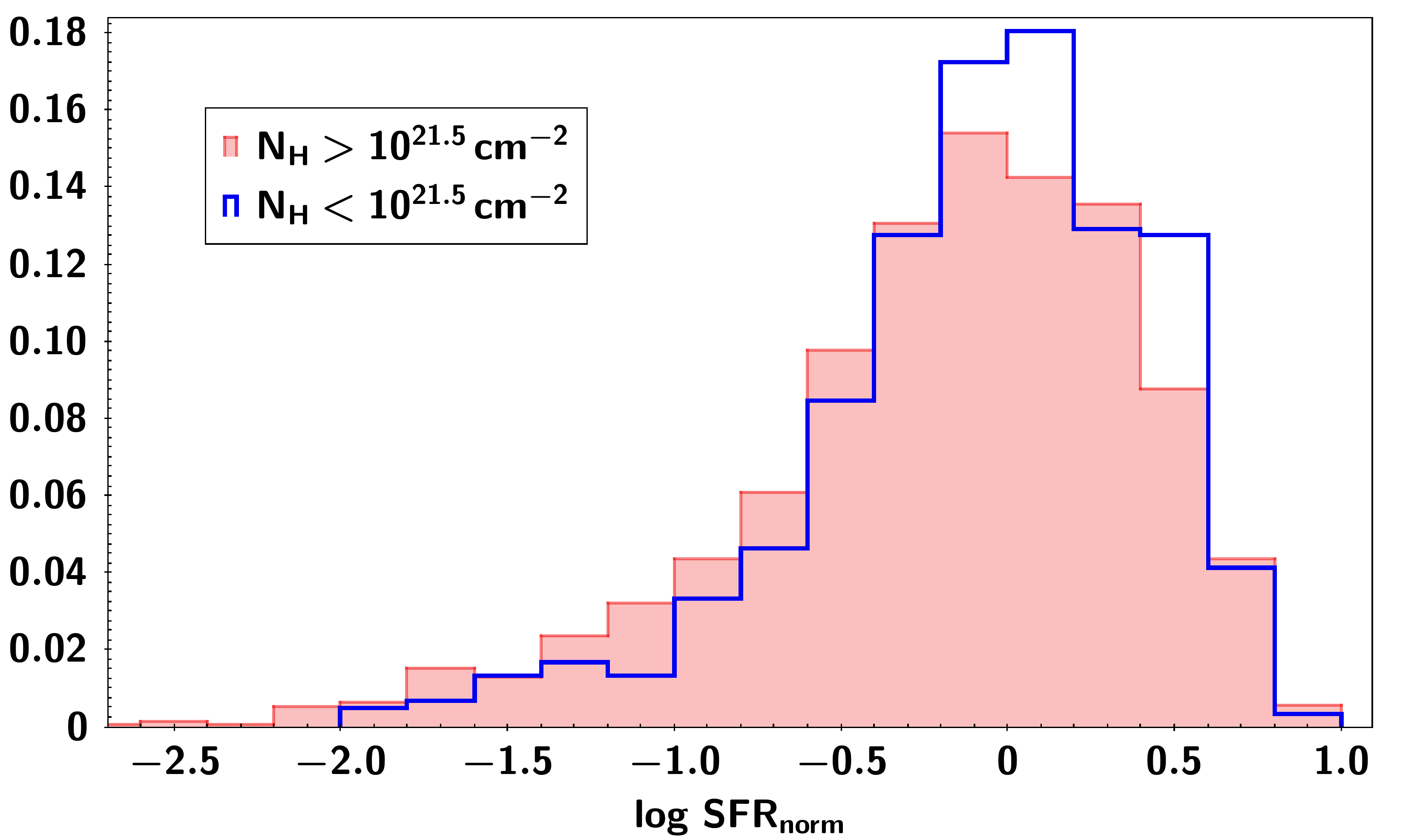}
%  \caption{}
  \label{fig_sfrnorm_xray}
\end{subfigure}
\caption{From top to bottom: Stellar mass, SFR and SFR$_{norm}$ distributions of X-ray absorbed (red, shaded histogram) and unabsorbed (blue line) AGN. The two populations have similar host galaxy properties.}
\label{fig_xray_host_properties}
\end{figure}

A plausible process that may link star formation with AGN activity, are galaxy mergers in gas rich galactic disks \citep[e.g.][]{Hopkins2008a}. This AGN fuelling mechanism is associated with star formation events and assumes that a fraction of the cold gas in galaxies that is available for star formation, accretes onto the SMBH \citep[e.g.][]{Bower2006, Fanidakis2012}. In higher mass systems, other mechanisms may fuel the SMBH, that are decoupled from the star formation of the host galaxy. For example, the SMBH may be activated when diffuse hot gas in quasi-hydrostatic equilibrium accretes onto the SMBH without first being cooled onto the galactic disk. Based on semi-analytic models, this mechanism becomes dominant at very massive systems \citep[e.g.][]{Fanidakis2013}. The above scenario implies that, if the correlation between AGN activity and SFR exists, this correlation is the result of a common mechanism that affects both properties. In other words, there is not an actual connection between the two parameters. 

%However, there may be a connection and not only correlation between AGN and SF. \cite{Zubovas2013} found that in gas rich phases, AGN outflows may trigger star formation by over-compress cold dense gas and provide positive feedback on the host galaxy. Star formation feedback could counteract this AGN outflow compression. Thus, the flat relation we observe could be the net result of two opposite acting processes. 

We conclude, that our results may suggest that SFR of luminous X-ray AGN is enhanced compared to SFR of star forming galaxies, in less massive systems (($\rm log\,[M_*(M_\odot)] \sim 11$)). However, our results are only tentative and further investigation is required before strong conclusions can be made. Additional data that would be mass-complete to lower masses could provide stronger evidence.

\section{Results II: The role of absorption on host galaxy properties}
\label{sec_host_properties}

In this Section, we examine whether absorbed X-ray AGN present different host galaxy properties compared to their unabsorbed counterparts. Specifically, we compare the stellar mass, SFR and SFR$_{norm}$ of the two populations. In this part of our analysis, we use the full X-ray catalogue, i.e., the 1989 X-ray AGN (see Section \ref{sec_sample_selection}). 

%Previous studies have found that optical/mid-IR colours and X-ray criteria produce different absorbed AGN samples \citep[e.g.][]{Ruiz2021, Mountrichas2020} since the correlation between optical/IR obscuration and X-ray absorption presents large scatter \citep[e.g.][]{Jaffarian2020}. This scatter is attributed to various reasons, such as X-ray variability \citep[e.g.][]{Yang2016}, absorbing material located at galactic scales \citep[e.g.][]{Malizia2020} and that different obscuration criteria are sensitive to different amounts of absorption \citep[e.g.][]{Masoura2020}, among others. Therefore, in our analysis we split the X-ray sources into absorbed and unabsorbed systems, using both X-ray and optical/mid-IR colour criteria.

\subsection{Classification based on X-ray obscuration}
We use the $\rm N_H$ parameter to classify sources into X-ray absorbed and unabsorbed, using a cut at $\rm N_H=10^{21.5}\,cm^{-2}$. We choose this threshold as it provides a good agreement between X-ray and optical classification of type 1 and 2 AGN \citep[e.g.][]{Merloni2014}. Using a higher $\rm N_H$ cut ($\rm N_H=10^{22}\,cm^{-2}$) does not change our results and conclusions. There are 374 (550) X-ray unabsorbed and 965 (778) absorbed AGN, using $\rm N_H=10^{21.5}\,cm^{-2}$ ($\rm =10^{22}\,cm^{-2}$). $\rm N_H$ measurements become less accurate for sources with small number of counts (photons). Moreover, $\rm N_H$ values, derived by HR measurements, are less secure, as we move to higher redshifts. This is because the absorption redshifts out of the soft X-ray band in the observed frame. Restricting our sample to sources with 50 or more net counts and at $\rm z<1$ does not change our conclusions.

Fig. \ref{fig_z_Lx_xray} presents the redshift and X-ray luminosity distributions of the two populations. The distributions of absorbed and unabsorbed sources are similar. Nevertheless, we account for the small differences. For that purpose, we join the redshift distributions of the two populations (and similarly the L$_X$ distributions) and normalize them to the total number of sources in each redshift (L$_X$) bin. This gives us the PDF in this 2$-$D (z, L$_X$) parameter space. Then, each source is weighted, based on its redshift and X-ray luminosity, according to the estimated PDF \citep{Mendez2016, Mountrichas2016}.

The top panel of Fig. \ref{fig_xray_host_properties}, presents the M$_*$ distributions of X-ray absorbed and unabsorbed X-ray AGN, in bins of 0.2\,dex. The two distributions are similar, as confirmed by the two-sample Kolmogorov-Smirnov test (KS$-$test, $p-\rm value = 0.12$). Our results are in agreement with most studies that used X-ray criteria for the classification of X-ray AGN \citep[][]{Merloni2014, Masoura2021}. However, \cite{Lanzuisi2017} found that N$_H$ and M$_*$ show a clear positive correlation, using  $\rm N_H=10^{22}\,cm^{-2}$ to classify their sources. Adopting the same N$_H$ threshold does not change our results. \cite{Zou2019}, used optical spectra, morphology and variability to classify 2463 X-ray selected AGN in the COSMOS field. Their analysis showed that type 1 AGN tend to have lower M$_*$ than type 2. The disagreement with our results could be attributed to the different classification criteria. 

In the middle panel of Fig. \ref{fig_xray_host_properties}, we present the SFR distributions of the two X-ray populations, in bins of 0.2\,dex. There is no significant difference between the SFR of X-ray absorbed and unabsorbed sources ($p-\rm value = 0.33$ from KS$-$test). Our results are in agreement with previous X-ray studies \citep{Rosario2012, Merloni2014, Lanzuisi2017, Zou2019, Masoura2021}.

The bottom panel of Fig. \ref{fig_xray_host_properties}, presents the SFR$_{norm}$ distributions of X-ray absorbed and unabsorbed AGN, in bins of 0.2\,dex. SFR$_{norm}$ distributions of the two AGN populations are similar ($p-\rm value = 0.27$ from KS$-$test). Our findings are in agreement with \cite{Masoura2021}.

Based on our results, X-ray absorbed and unabsorbed AGN reside in galaxies with similar properties.

\section{Summary-Conclusions}

In this work, we used X-ray sources observed by the {\it{Chandra}} X-ray Observatory within the 9.3\,deg$^2$ Bo$\rm \ddot{o}$tes field of the NDWFS \citep[][catalogue]{Masini2020} to study whether there is a link between the AGN power and the star formation of the host galaxy, at $\rm 0.5<z<2.0$. After applying our selection criteria (e.g. photometry, mass completeness; Section \ref{sec_sample_selection}), the X-ray sample consists of 711 sources (Table \ref{table_data}). About half of these sources have spectroscopic redshifts and a similar fraction has been observed by {\it{Herschel}}. Furthermore, a reference galaxy catalogue is constructed with which the SFR of the X-ray sources are compared. The same selection criteria are applied in the reference sample. Additionally, sources with X-ray emission and a strong AGN component are excluded from the latter dataset. The reference catalogue numbers 11,639 galaxies. 

For both datasets, we use photometric data compiled by the HELP project and apply SED fitting using the X-CIGALE code. This enables us measure the properties of the sources (e.g. SFR, stellar mass). In the fitting process, we include the X-ray flux that is available in the X-ray catalogue and account for possible presence of a polar dust component.

We study the SFR-L$_X$ relation with respect to the position of the galaxy to the MS. For that purpose, we estimate the SFR$_{norm}$ parameter. We use SFR measurements from X-CIGALE, for the X-ray and the reference galaxy catalogues. Quiescent systems are excluded from both datasets. From the galaxy reference catalogue, we also reject sources with a strong AGN component ($\rm frac_{AGN}>0.2$). We detect only a mild correlation between SFR$_{norm}$ and L$_X$. Specifically, SFR$_{norm}$ increases by $20-30\%$ up to $\rm L_{X,2-10keV}\sim10^{44.5}\,ergs^{-1}$. 

SFR$_{norm}$ is also estimated using the SFR measurements of X-CIGALE for the X-ray sources while for the SFR of star forming galaxies, we use equation 9 from \cite{Schreiber2015}. This allows us to compare our measurements with those from the literature that have used the same approach. In this case, SFR$_{norm}$ increases by a factor of 2, within an order of magnitude in L$_X$. We do not detect evolution of SFR$_{norm}$ with redshift, within $\rm 0.5<z<2.5$.

These results highlight the importance of performing a uniform and consistent analysis when studying the correlation between the SFR of a galaxy with the X-ray luminosity. Systematics that are inserted in the analysis due to the different methodologies used in the estimation of SFR of AGN and non AGN systems among different studies, the different definitions of the star forming MS among them and not accounting for the mass incompleteness of the samples at different redshift, may greatly affect the measurements and lead to incorrect conclusions.

Using our SFR measurements for both X-ray and star forming galaxies, we study the dependence of SFR$_{norm}$ on the specific accretion rate. SFR$_{norm}$ increases by $\sim 20\%$  within the $\lambda _{s\,BHAR}$ range spanned by the dataset. Prompted by this result, we split our samples into stellar mass bins and revisit the $\rm SFR_{norm}-L_X$ relation. Our analysis suggests that, in less massive systems ($\rm log\,[M_*(M_\odot)] \sim 11$), the SFR of X-ray AGN is enhanced compared to that of non X-ray AGN galaxies by $\sim 50\%$. In the most massive galaxies, a flat relation is detected. Our results, although tentative, are consistent with a scenario in which mergers trigger the AGN activity and the star formation of the host galaxy, by increasing the available cold gas, while in the most massive systems ($\rm log\,[M_*(M_\odot)] > 11.5$), other mechanisms, that are decoupled from the star formation, fuel the SMBH (e.g. diffuse hot gas).

Finally, we split the X-ray sample into X-ray obscured and unobscured AGN, by applying a cut at $\rm N_H=10^{21.5}\,cm^{-2}$ and examine whether the host galaxy properties of the two populations differ. Our analysis, showed that both AGN types have similar SFR, stellar mass and SFR$_{norm}$ distributions. This suggests that X-ray absorption is not linked with the properties of the host galaxy. 
 
Our analysis and results highlight the fact that AGN constitute a diverse population of galaxies with a wide range of  e.g. SMBH fuelling mechanisms, luminosities, stellar masses, SFR and across large cosmological epochs. Therefore, to accurately measure the effect of one parameter on to another we need first to disentangle all other parameters from the analysis. This is extremely challenging taking into account the number of available X-ray sources and the selection biases that affect the datasets and thus the final results. Ongoing and future X-ray surveys (eROSITA, {\it{Athena}}) will provide us with large samples of X-ray sources, orders of magnitude larger than current datasets. Along with consistent methodologies and improved machinery these X-ray surveys will enable us to shed light on galaxy evolution and the interplay between SMBH and their host galaxies.

\begin{acknowledgements}
The authors thank the anonymous referee for their detailed report that improved the quality of the paper.
\\
GM acknowledges support by the Agencia Estatal de Investigación, Unidad de Excelencia María de Maeztu, ref. MDM-2017-0765.
\\
MB acknowledges FONDECYT regular grant 1170618
\\
The project has received funding from Excellence Initiative of Aix-Marseille University - AMIDEX, a French `Investissements d'Avenir' programme.
\\
KM has been supported by the National Science Centre (UMO-2018/30/E/ST9/00082).

\end{acknowledgements}

\bibliography{mybib}{}

\begin{thebibliography}{85}
\expandafter\ifx\csname natexlab\endcsname\relax\def\natexlab#1{#1}\fi

\bibitem[{Aird {et~al.}(2018)Aird, Coil, \& Georgakakis}]{Aird2018}
Aird, J., Coil, A.~L., \& Georgakakis, A. 2018, Monthly Notices of the Royal
  Astronomical Society, 474, 1225

\bibitem[{Aird {et~al.}(2019)Aird, Coil, \& Georgakakis}]{Aird2019}
Aird, J., Coil, A.~L., \& Georgakakis, A. 2019, Monthly Notices of the Royal
  Astronomical Society, 484, 4360

\bibitem[{Aird {et~al.}(2012)Aird, Coil, Moustakas, Blanton, Burles, Cool,
  Eisenstein, Smith, Wong, \& Zhu}]{Aird2012}
Aird, J., Coil, A.~L., Moustakas, J., {et~al.} 2012, The Astrophysical Journal,
  746, 90

\bibitem[{{Alexander} \& {Hickox}(2012)}]{Alexander2012}
{Alexander}, D.~M. \& {Hickox}, R.~C. 2012, NewAR, 56, 93

\bibitem[{{Allevato} {et~al.}(2011)}]{Allevato2011}
{Allevato}, V. {et~al.} 2011, ApJ, 736, 99

\bibitem[{Antonucci(1993)}]{Antonucci1993}
Antonucci, R. 1993, Annual Review of Astronomy and Astrophysics, 31, 473

\bibitem[{Bernhard {et~al.}(2019)Bernhard, Grimmett, Mullaney, Daddi,
  Tadhunter, \& Jin}]{Bernhard2019}
Bernhard, E., Grimmett, L.~P., Mullaney, J.~R., {et~al.} 2019, Monthly Notices
  of the Royal Astronomical Society: Letters, 483, L52

\bibitem[{Bezanson {et~al.}(2012)Bezanson, van Dokkum, \& Franx}]{Bezanson2012}
Bezanson, R., van Dokkum, P., \& Franx, M. 2012, The Astrophysical Journal,
  760, 62

\bibitem[{Blandford \& Rees(1974)}]{Blandford1974}
Blandford, R.~D. \& Rees, M.~J. 1974, Monthly Notices of the Royal Astronomical
  Society, 169, 395

\bibitem[{Boquien {et~al.}(2019)Boquien, Burgarella, Roehlly, Buat, Ciesla,
  Corre, Inoue, \& Salas}]{Boquien2019}
Boquien, M., Burgarella, D., Roehlly, Y., {et~al.} 2019, Astronomy {\&}
  Astrophysics, 622, A103

\bibitem[{{Bower} {et~al.}(2006){Bower}, {Benson}, {Malbon}, {Helly}, {Frenk},
  {Baugh}, {Cole}, \& {Lacey}}]{Bower2006}
{Bower}, R.~G., {Benson}, A.~J., {Malbon}, R., {et~al.} 2006, MNRAS, 370, 645

\bibitem[{Boyle {et~al.}(2000)Boyle, Shanks, Croom, Smith, Miller, Loaring, \&
  Heymans}]{Boyle2000}
Boyle, B.~J., Shanks, T., Croom, S.~M., {et~al.} 2000, Monthly Notices of the
  Royal Astronomical Society, 317, 1014

\bibitem[{Boyle \& Terlevich(1998)}]{Boyle1998}
Boyle, B.~J. \& Terlevich, R.~J. 1998, MNRAS, 293, 49

\bibitem[{Brown {et~al.}(2019)Brown, Nayyeri, Cooray, Ma, Hickox, \&
  Azadi}]{Brown2019}
Brown, A., Nayyeri, H., Cooray, A., {et~al.} 2019, The Astrophysical Journal,
  871, 87

\bibitem[{Bruzual \& Charlot(2003)}]{Bruzual_Charlot2003}
Bruzual, G. \& Charlot, S. 2003, MNRAS, 344, 1000

\bibitem[{Buat {et~al.}(2019)Buat, Ciesla, Boquien, Ma{\l}ek, \&
  Burgarella}]{Buat2019}
Buat, V., Ciesla, L., Boquien, M., Ma{\l}ek, K., \& Burgarella, D. 2019,
  Astronomy {\&} Astrophysics, 632, A79

\bibitem[{Chabrier(2003)}]{Chabrier2003}
Chabrier, G. 2003, PASP, 115, 763

\bibitem[{Charlot \& Fall(2000)}]{Charlot_Fall_2000}
Charlot, S. \& Fall, S.~M. 2000, ApJ, 539, 718

\bibitem[{Chen {et~al.}(2015)Chen, Hickox, Alberts, Harrison, Alexander, Assef,
  Brodwin, Brown, Moro, Forman, Gorjian, Goulding, Hainline, Jones, Kochanek,
  Murray, Pope, Rovilos, \& Stern}]{Chen2015}
Chen, C.-T.~J., Hickox, R.~C., Alberts, S., {et~al.} 2015, The Astrophysical
  Journal, 802, 50

\bibitem[{Ciotti \& Ostriker(1997)}]{Ciotti1997}
Ciotti, L. \& Ostriker, J.~P. 1997, The Astrophysical Journal, 487, L105

\bibitem[{Circosta {et~al.}(2019)Circosta, Vignali, Gilli, Feltre, Vito,
  Calura, Mainieri, Massardi, \& Norman}]{Circosta2019}
Circosta, C., Vignali, C., Gilli, R., {et~al.} 2019, Astronomy {\&}
  Astrophysics, 623, A172

\bibitem[{{Dale} {et~al.}(2014){Dale}, {Helou}, {Magdis}, {Armus},
  {D{\'{\i}}az-Santos}, \& {Shi}}]{Dale2014}
{Dale}, D.~A., {Helou}, G., {Magdis}, G.~E., {et~al.} 2014, ApJ, 784, 83

\bibitem[{DeBuhr {et~al.}(2012)DeBuhr, Quataert, \& Ma}]{Debuhr2012}
DeBuhr, J., Quataert, E., \& Ma, C.-P. 2012, Monthly Notices of the Royal
  Astronomical Society, 420, 2221

\bibitem[{Duncan {et~al.}(2018{\natexlab{a}})Duncan, Brown, Williams, Best,
  Buat, Burgarella, Jarvis, Ma{\l}ek, Oliver, Röttgering, \&
  Smith}]{Duncan2018a}
Duncan, K.~J., Brown, M. J.~I., Williams, W.~L., {et~al.} 2018{\natexlab{a}},
  Monthly Notices of the Royal Astronomical Society, 473, 2655

\bibitem[{Duncan {et~al.}(2018{\natexlab{b}})Duncan, Jarvis, Brown, \&
  Röttgering}]{Duncan2018b}
Duncan, K.~J., Jarvis, M.~J., Brown, M. J.~I., \& Röttgering, H. J.~A.
  2018{\natexlab{b}}, Monthly Notices of the Royal Astronomical Society

\bibitem[{Duncan {et~al.}(2019)Duncan, Sabater, Röttgering, Jarvis, Smith,
  Best, Callingham, Cochrane, Croston, Hardcastle, Mingo, Morabito, Nisbet,
  Prandoni, Shimwell, Tasse, White, Williams, Alegre, Chy{\.{z}}y, Gürkan,
  Hoeft, Kondapally, Mechev, Miley, Schwarz, \& van Weeren}]{Duncan2019}
Duncan, K.~J., Sabater, J., Röttgering, H. J.~A., {et~al.} 2019, Astronomy
  {\&} Astrophysics, 622, A3

\bibitem[{Eales {et~al.}(2010)Eales, Dunne, Clements, Cooray, De~Zotti, Dye,
  Ivison, Jarvis, Lagache, Maddox, Negrello, Serjeant, Thompson, Van~Kampen,
  Amblard, Andreani, Baes, Beelen, Bendo, Benford, Bertoldi, Bock, Bonfield,
  Boselli, Bridge, Buat, Burgarella, Carlberg, Cava, Chanial, Charlot,
  Christopher, Coles, Cortese, Dariush, da~Cunha, Dalton, Danese, Dannerbauer,
  Driver, Dunlop, Fan, Farrah, Frayer, Frenk, Geach, Gardner, Gomez,
  Gonz{\'a}lez-Nuevo, Gonz{\'a}lez-Solares, Griffin, Hardcastle,
  Hatziminaoglou, Herranz, Hughes, Ibar, Jeong, Lacey, Lapi, Lawrence, Lee,
  Leeuw, Liske, L{\'o}pez-Caniego, M{\"u}ller, Nandra, Panuzzo, Papageorgiou,
  Patanchon, Peacock, Pearson, Phillipps, Pohlen, Popescu, Rawlings, Rigby,
  Rigopoulou, Robotham, Rodighiero, Sansom, Schulz, Scott, Smith, Sibthorpe,
  Smail, Stevens, Sutherland, Takeuchi, Tedds, Temi, Tuffs, Trichas, Vaccari,
  Valtchanov, van~der Werf, Verma, Vieria, Vlahakis, \& White}]{Eales2010}
Eales, S., Dunne, L., Clements, D., {et~al.} 2010, PASP, 122, 499

\bibitem[{Elbaz {et~al.}(2007)Elbaz, Daddi, Borgne, Dickinson, Alexander,
  Chary, Starck, Brandt, Kitzbichler, MacDonald, Nonino, Popesso, Stern, \&
  Vanzella}]{Elbaz2007}
Elbaz, D., Daddi, E., Borgne, D.~L., {et~al.} 2007, Astronomy {\&}
  Astrophysics, 468, 33

\bibitem[{Fanidakis {et~al.}(2013)Fanidakis, Georgakakis, Mountrichas, Krumpe,
  Baugh, Lacey, Frenk, Miyaji, \& Benson}]{Fanidakis2013}
Fanidakis, N., Georgakakis, A., Mountrichas, G., {et~al.} 2013, Monthly Notices
  of the Royal Astronomical Society, 435, 679

\bibitem[{{Fanidakis} {et~al.}(2012)}]{Fanidakis2012}
{Fanidakis}, N. {et~al.} 2012, MNRAS, 419, 2797

\bibitem[{{Ferrarese} \& {Merritt}(2000)}]{Ferrarese2000}
{Ferrarese}, L. \& {Merritt}, D. 2000, ApJ, 539, 9

\bibitem[{Florez {et~al.}(2020)Florez, Jogee, Sherman, Stevans, Finkelstein,
  Papovich, Kawinwanichakij, Ciardullo, Gronwall, Urry, Kirkpatrick, LaMassa,
  Ananna, \& Wold}]{Florez2020}
Florez, J., Jogee, S., Sherman, S., {et~al.} 2020, Monthly Notices of the Royal
  Astronomical Society, 497, 3273

\bibitem[{Genzel {et~al.}(2014)Genzel, Schreiber, Rosario, Lang, Lutz,
  Wisnioski, Wuyts, Wuyts, Bandara, Bender, Berta, Kurk, Mendel, Tacconi,
  Wilman, Beifiori, Brammer, Burkert, Buschkamp, Chan, Carollo, Davies,
  Eisenhauer, Fabricius, Fossati, Kriek, Kulkarni, Lilly, Mancini, Momcheva,
  Naab, Nelson, Renzini, Saglia, Sharples, Sternberg, Tacchella, \& van
  Dokkum}]{Genzel2014}
Genzel, R., Schreiber, N. M.~F., Rosario, D., {et~al.} 2014, The Astrophysical
  Journal, 796, 7

\bibitem[{Georgakakis {et~al.}(2017)Georgakakis, Aird, Schulze, Dwelly,
  Salvato, Nandra, Merloni, \& Schneider}]{Georgakakis2017b}
Georgakakis, A., Aird, J., Schulze, A., {et~al.} 2017, MNRAS, 471, 1976

\bibitem[{Grimmett {et~al.}(2020)Grimmett, Mullaney, Bernhard, Harrison,
  Alexander, Stanley, Masoura, \& Walters}]{Grimmett2020}
Grimmett, L.~P., Mullaney, J.~R., Bernhard, E.~P., {et~al.} 2020, Monthly
  Notices of the Royal Astronomical Society, 495, 1392

\bibitem[{{Harrison} {et~al.}(2012)}]{Harrison2012}
{Harrison}, C.~M. {et~al.} 2012, ApJL, 760, 5

\bibitem[{Hickox {et~al.}(2014)Hickox, Mullaney, Alexander, Chen, Civano, \&
  Goulding}]{Hickox2014}
Hickox, R.~C., Mullaney, J.~R., Alexander, D.~M., {et~al.} 2014, ApJ, 782, 11

\bibitem[{{Hopkins} {et~al.}(2008){Hopkins}, {Hernquist}, {Cox}, \&
  {Keres}}]{Hopkins2008a}
{Hopkins}, P.~F., {Hernquist}, L., {Cox}, T.~J., \& {Keres}, D. 2008, ApJS,
  175, 356

\bibitem[{Hopkins {et~al.}(2006)Hopkins, Hernquist, Cox, Matteo, Robertson, \&
  Springel}]{Hopkins2006}
Hopkins, P.~F., Hernquist, L., Cox, T.~J., {et~al.} 2006, The Astrophysical
  Journal Supplement Series, 163, 1

\bibitem[{Hurley {et~al.}(2017)Hurley, Oliver, Betancourt, Clarke, Cowley,
  Duivenvoorden, Farrah, Griffin, Lacey, Floch, Papadopoulos, Sargent, Scudder,
  Vaccari, Valtchanov, \& Wang}]{Hurley2017}
Hurley, P.~D., Oliver, S., Betancourt, M., {et~al.} 2017, Monthly Notices of
  the Royal Astronomical Society, 464, 885

\bibitem[{{Just} {et~al.}(2007){Just}, {Brandt}, {Shemmer}, {Steffen},
  {Schneider}, {Chartas}, \& {Garmire}}]{Just2007}
{Just}, D.~W., {Brandt}, W.~N., {Shemmer}, O., {et~al.} 2007, ApJ, 685, 1004

\bibitem[{Kochanek {et~al.}(2012)Kochanek, Eisenstein, Cool, Caldwell, Assef,
  Jannuzi, Jones, Murray, Forman, Dey, Brown, Eisenhardt, Gonzalez, Green, \&
  Stern}]{Kochanek2012}
Kochanek, C.~S., Eisenstein, D.~J., Cool, R.~J., {et~al.} 2012, The
  Astrophysical Journal Supplement Series, 200, 8

\bibitem[{{Lanzuisi} {et~al.}(2017)}]{Lanzuisi2017}
{Lanzuisi}, G. {et~al.} 2017, A\&A, 602, 13

\bibitem[{Li {et~al.}(2019)Li, Xue, Sun, Liu, Vito, Brandt, Hughes, Yang,
  Tozzi, Zhu, Zheng, Luo, Chen, Vignali, Gilli, \& Shu}]{Li2019}
Li, J., Xue, Y., Sun, M., {et~al.} 2019, The Astrophysical Journal, 877, 5

\bibitem[{Loh(2008)}]{Loh2008}
Loh, J.~M. 2008, ApJ, 681, 726

\bibitem[{{Lutz} {et~al.}(2010)}]{Lutz2010}
{Lutz}, D. {et~al.} 2010, ApJ, 712, 1287

\bibitem[{Magorrian {et~al.}(1998)}]{Magorrian1998}
Magorrian, J. {et~al.} 1998, AJ, 115, 2285

\bibitem[{Maiolino \& Rieke(1995)}]{Maiolino1995}
Maiolino, R. \& Rieke, G.~H. 1995, The Astrophysical Journal, 454, 95

\bibitem[{Malizia {et~al.}(2020)Malizia, Bassani, Stephen, Bazzano, \&
  Ubertini}]{Malizia2020}
Malizia, A., Bassani, L., Stephen, J.~B., Bazzano, A., \& Ubertini, P. 2020,
  Astronomy {\&} Astrophysics, 639, A5

\bibitem[{Masini {et~al.}(2020)Masini, Hickox, Carroll, Aird, Alexander, Assef,
  Bower, Brodwin, Brown, Chatterjee, Chen, Dey, DiPompeo, Duncan, Eisenhardt,
  Forman, Gonzalez, Goulding, Hainline, Jannuzi, Jones, Kochanek, Kraft, Lee,
  Miller, Mullaney, Myers, Ptak, Stanford, Stern, Vikhlinin, Wake, \&
  Murray}]{Masini2020}
Masini, A., Hickox, R.~C., Carroll, C.~M., {et~al.} 2020, The Astrophysical
  Journal Supplement Series, 251, 2

\bibitem[{Masoura {et~al.}(2020)Masoura, Georgantopoulos, Mountrichas, Vignali,
  Koulouridis, Chiappetti, Fotopoulou, Paltani, \& Pierre}]{Masoura2020}
Masoura, V.~A., Georgantopoulos, I., Mountrichas, G., {et~al.} 2020, Astronomy
  {\&} Astrophysics, 638, A45

\bibitem[{Masoura {et~al.}(2021)Masoura, Mountrichas, Georgantopoulos, \&
  Plionis}]{Masoura2021}
Masoura, V.~A., Mountrichas, G., Georgantopoulos, I., \& Plionis, M. 2021,
  Astronomy {\&} Astrophysics, 646, A167

\bibitem[{Masoura {et~al.}(2018)Masoura, Mountrichas, Georgantopoulos, Ruiz,
  Magdis, \& Plionis}]{Masoura2018}
Masoura, V.~A., Mountrichas, G., Georgantopoulos, I., {et~al.} 2018, A\&A, 618,
  31

\bibitem[{{Mendez} {et~al.}(2016)}]{Mendez2016}
{Mendez}, A.~J. {et~al.} 2016, ApJ, 821, 55

\bibitem[{Merloni {et~al.}(2014)Merloni, Bongiorno, Brusa, Iwasawa, Mainieri,
  Magnelli, Salvato, Berta, Cappelluti, Comastri, Fiore, Gilli, \&
  Koekemoer}]{Merloni2014}
Merloni, A., Bongiorno, A., Brusa, M., {et~al.} 2014, Monthly Notices of the
  Royal Astronomical Society, 437, 3550

\bibitem[{Mountrichas {et~al.}(2021)Mountrichas, Buat, Yang, Boquien,
  Burgarella, \& Ciesla}]{Mountrichas2021}
Mountrichas, G., Buat, V., Yang, G., {et~al.} 2021, Astronomy {\&}
  Astrophysics, 646, A29

\bibitem[{Mountrichas {et~al.}(2013)}]{Mountrichas2013}
Mountrichas, G. {et~al.} 2013, MNRAS, 430, 661

\bibitem[{{Mountrichas} {et~al.}(2016)}]{Mountrichas2016}
{Mountrichas}, G. {et~al.} 2016, MNRAS, 457, 4195

\bibitem[{Mullaney {et~al.}(2015)Mullaney, Alexander, Aird, Bernhard, Daddi,
  Moro, Dickinson, Elbaz, Harrison, Juneau, Liu, Pannella, Rosario, Santini,
  Sargent, Schreiber, Simpson, \& Stanley}]{Mullaney2015}
Mullaney, J.~R., Alexander, D.~M., Aird, J., {et~al.} 2015, Monthly Notices of
  the Royal Astronomical Society: Letters, 453, L83

\bibitem[{Noeske {et~al.}(2007)Noeske, Weiner, Faber, Papovich, Koo,
  Somerville, Bundy, Conselice, Newman, Schiminovich, Floch, Coil, Rieke, Lotz,
  Primack, Barmby, Cooper, Davis, Ellis, Fazio, Guhathakurta, Huang, Kassin,
  Martin, Phillips, Rich, Small, Willmer, \& Wilson}]{Noeske2007}
Noeske, K.~G., Weiner, B.~J., Faber, S.~M., {et~al.} 2007, The Astrophysical
  Journal, 660, L43

\bibitem[{Ogawa {et~al.}(2021)Ogawa, Ueda, Tanimoto, \& Yamada}]{Ogawa2021}
Ogawa, S., Ueda, Y., Tanimoto, A., \& Yamada, S. 2021, The Astrophysical
  Journal, 906, 84

\bibitem[{Oliver {et~al.}(2012)Oliver, Bock, Altieri, Amblard, Arumugam,
  Aussel, Babbedge, Beelen, B{\'{e}}thermin, Blain, Boselli, Bridge, Brisbin,
  Buat, Burgarella, Castro-Rodr{\'{\i}}guez, Cava, Chanial, Cirasuolo,
  Clements, Conley, Conversi, Cooray, Dowell, Dubois, Dwek, Dye, Eales, Elbaz,
  Farrah, Feltre, Ferrero, Fiolet, Fox, Franceschini, Gear, Giovannoli, Glenn,
  Gong, Solares, Griffin, Halpern, Harwit, Hatziminaoglou, Heinis, Hurley,
  Hwang, Hyde, Ibar, Ilbert, Isaak, Ivison, Lagache, Floch, Levenson, Faro, Lu,
  Madden, Maffei, Magdis, Mainetti, Marchetti, Marsden, Marshall, Mortier,
  Nguyen, Halloran, Omont, Page, Panuzzo, Papageorgiou, Patel, Pearson,
  Perez-Fournon, Pohlen, Rawlings, Raymond, Rigopoulou, Riguccini, Rizzo,
  Rodighiero, Roseboom, Rowan-Robinson, Portal, Schulz, Scott, Seymour, Shupe,
  Smith, Stevens, Symeonidis, Trichas, Tugwell, Vaccari, Valtchanov, Vieira,
  Viero, Vigroux, Wang, Ward, Wardlow, Wright, Xu, \& Zemcov}]{Oliver2012a}
Oliver, S.~J., Bock, J., Altieri, B., {et~al.} 2012, Monthly Notices of the
  Royal Astronomical Society, 424, 1614

\bibitem[{{Oliver} {et~al.}(2012)}]{Oliver2012}
{Oliver}, S.~J. {et~al.} 2012, MNRAS, 424, 1614

\bibitem[{{Page} {et~al.}(2012)}]{Page2012}
{Page}, M.~J. {et~al.} 2012, nat, 485, 213

\bibitem[{Pannella {et~al.}(2009)Pannella, Carilli, Daddi, McCracken, Owen,
  Renzini, Strazzullo, Civano, Koekemoer, Schinnerer, Scoville,
  Smol{\v{c}}i{\'{c}}, Taniguchi, Aussel, Kneib, Ilbert, Mellier, Salvato,
  Thompson, \& Willott}]{Pannella2009}
Pannella, M., Carilli, C.~L., Daddi, E., {et~al.} 2009, The Astrophysical
  Journal, 698, L116

\bibitem[{Park {et~al.}(2006)Park, Kashyap, Siemiginowska, van Dyk, Zezas,
  Heinke, \& Wargelin}]{Park2006}
Park, T., Kashyap, V.~L., Siemiginowska, A., {et~al.} 2006, The Astrophysical
  Journal, 652, 610

\bibitem[{{Pozzetti} {et~al.}(2010)}]{Pozzetti2010}
{Pozzetti}, L. {et~al.} 2010, A\&A, 523, 23

\bibitem[{Prevot {et~al.}(1984)Prevot, Lequeux, Maurice, Prevot, \&
  Rocca-Volmerange}]{Prevot1984}
Prevot, M., Lequeux, J., Maurice, E., Prevot, L., \& Rocca-Volmerange, B. 1984,
  A\&A, 132, 389

\bibitem[{Risaliti \& Lusso(2017)}]{Risaliti2017}
Risaliti, G. \& Lusso, E. 2017, Astronomische Nachrichten, 338, 329

\bibitem[{Rosario {et~al.}(2013)Rosario, Trakhtenbrot, Lutz, Netzer, Trump,
  Silverman, Schramm, Lusso, Berta, Bongiorno, Brusa, Förster-Schreiber,
  Genzel, Lilly, Magnelli, Mainieri, Maiolino, Merloni, Mignoli, Nordon,
  Popesso, Salvato, Santini, Tacconi, \& Zamorani}]{Rosario2013}
Rosario, D.~J., Trakhtenbrot, B., Lutz, D., {et~al.} 2013, Astronomy {\&}
  Astrophysics, 560, A72

\bibitem[{Rosario {et~al.}(2012)}]{Rosario2012}
Rosario, D.~J. {et~al.} 2012, A\&A, 545, 18

\bibitem[{Santini {et~al.}(2012)Santini, Rosario, Shao, Lutz, Maiolino,
  Alexander, Altieri, Andreani, Aussel, Bauer, Berta, Bongiovanni, Brandt,
  Brusa, Cepa, Cimatti, Daddi, Elbaz, Fontana, Schreiber, Genzel, Grazian,
  Floc'h, Magnelli, Mainieri, Nordon, Garcia, Poglitsch, Popesso, Pozzi,
  Riguccini, Rodighiero, Salvato, Sanchez-Portal, Sturm, Tacconi, Valtchanov,
  \& Wuyts}]{Santini2012}
Santini, P., Rosario, D.~J., Shao, L., {et~al.} 2012, Astronomy {\&}
  Astrophysics, 540, A109

\bibitem[{Scheuer(1974)}]{Scheuer1974}
Scheuer, P. A.~G. 1974, Monthly Notices of the Royal Astronomical Society, 166,
  513

\bibitem[{Schreiber {et~al.}(2015)}]{Schreiber2015}
Schreiber, C. {et~al.} 2015, A\&A, 575, 29

\bibitem[{Shimizu {et~al.}(2015)Shimizu, Mushotzky, Mel{\'{e}}ndez, Koss, \&
  Rosario}]{Shimizu2015}
Shimizu, T.~T., Mushotzky, R.~F., Mel{\'{e}}ndez, M., Koss, M., \& Rosario,
  D.~J. 2015, Monthly Notices of the Royal Astronomical Society, 452, 1841

\bibitem[{Shirley {et~al.}(2019)Shirley, Roehlly, Hurley, Buat, del Carmen
  Campos~Varillas, Duivenvoorden, Duncan, Efstathiou, Farrah, Solares, Malek,
  Marchetti, McCheyne, Papadopoulos, Pons, Scipioni, Vaccari, \&
  Oliver}]{Shirley2019}
Shirley, R., Roehlly, Y., Hurley, P.~D., {et~al.} 2019, Monthly Notices of the
  Royal Astronomical Society, 490, 634

\bibitem[{Sobral {et~al.}(2013)Sobral, Smail, Best, Geach, Matsuda, Stott,
  Cirasuolo, \& Kurk}]{Sobral2013}
Sobral, D., Smail, I., Best, P.~N., {et~al.} 2013, Monthly Notices of the Royal
  Astronomical Society, 428, 1128

\bibitem[{Stalevski {et~al.}(2012)Stalevski, Fritz, Baes, Nakos, \&
  Popovi{\'{c}}}]{Stalevski2012}
Stalevski, M., Fritz, J., Baes, M., Nakos, T., \& Popovi{\'{c}}, L.~{\v{C}}.
  2012, Monthly Notices of the Royal Astronomical Society, 420, 2756

\bibitem[{Stalevski {et~al.}(2016)Stalevski, Ricci, Ueda, Lira, Fritz, \&
  Baes}]{Stalevski2016}
Stalevski, M., Ricci, C., Ueda, Y., {et~al.} 2016, Monthly Notices of the Royal
  Astronomical Society, 458, 2288

\bibitem[{Whitaker {et~al.}(2014)Whitaker, Franx, Leja, van Dokkum, Henry,
  Skelton, Fumagalli, Momcheva, Brammer, Labb{\'{e}}, Nelson, \&
  Rigby}]{Whitaker2014}
Whitaker, K.~E., Franx, M., Leja, J., {et~al.} 2014, The Astrophysical Journal,
  795, 104

\bibitem[{Yang {et~al.}(2020)Yang, Boquien, Buat, Burgarella, Ciesla, Duras,
  Stalevski, Brandt, \& Papovich}]{Yang2020}
Yang, G., Boquien, M., Buat, V., {et~al.} 2020, Monthly Notices of the Royal
  Astronomical Society, 491, 740

\bibitem[{Yang {et~al.}(2018)Yang, Brandt, Vito, Chen, Trump, Luo, Sun, Xue,
  Koekemoer, Schneider, Vignali, \& Wang}]{Yang2018}
Yang, G., Brandt, W.~N., Vito, F., {et~al.} 2018, Monthly Notices of the Royal
  Astronomical Society, 475, 1887

\bibitem[{Zinn {et~al.}(2013)Zinn, Middelberg, Norris, \& Dettmar}]{Zinn2013}
Zinn, P.-C., Middelberg, E., Norris, R.~P., \& Dettmar, R.-J. 2013, The
  Astrophysical Journal, 774, 66

\bibitem[{Zou {et~al.}(2019)Zou, Yang, Brandt, \& Xue}]{Zou2019}
Zou, F., Yang, G., Brandt, W.~N., \& Xue, Y. 2019, The Astrophysical Journal,
  878, 11

\bibitem[{Zubovas {et~al.}(2013)Zubovas, Nayakshin, King, \&
  Wilkinson}]{Zubovas2013}
Zubovas, K., Nayakshin, S., King, A., \& Wilkinson, M. 2013, Monthly Notices of
  the Royal Astronomical Society, 433, 3079

\end{thebibliography}
\bibliographystyle{aa}

\end{document}